# High-Resolution Rooftop-PV Potential Assessment for a Resilient Energy System in Ukraine

**Christoph Winkler**[1,2,†], **Kristina Dabrock**[1,2,†,*], **Serhiy Kapustyan**[1,†], **Craig Hart**[3], **Heidi Heinrichs**[1],

**Jann Michael Weinand**[1], **Jochen Linßen**[1], **and Detlef Stolten**[1,2]

[1] Forschungszentrum Jülich GmbH, Institute of Climate and Energy Systems, Jülich Systems

Analysis, 52425 Jülich, Germany

[2] RWTH Aachen University, Chair for Fuel Cells, Faculty of Mechanical Engineering, 52062 Aachen, Germany

[3] International Energy Agency (IEA), 75739 Paris, France

[†] Equally contributed

* Corresponding author: Kristina Dabrock, Forschungszentrum Jülich GmbH, Institute of Climate and Energy

Systems, Jülich Systems Analysis, 52425 Jülich, Germany; k.dabrock@fz-juelich.de

## CRediT authorship contribution statement

**Christoph Winkler**: Conceptualization; Data curation; Formal analysis; Investigation; Methodology; Software; Validation; Visualization; Writing - original draft; Writing – review & editing; Project administration

**Kristina Dabrock**: Conceptualization; Data curation; Formal analysis; Investigation; Methodology; Software; Validation; Visualization; Writing - original draft; Writing – review & editing

**Serhiy Kapustyan**: Conceptualization; Data curation; Investigation; Validation; Writing – original draft; Writing – review & editing

**Craig Hart**: Writing – review & editing

**Heidi Heinrichs**: Supervision, Writing – review & editing; Funding acquisition

**Jann Michael Weinand**: Supervision, Writing – review & editing

**Jochen Linßen**: Supervision; Resources; Funding acquisition

**Detlef Stolten**: Supervision; Resources; Funding acquisition

## Abstract

Rooftop photovoltaic (RTPV) systems are essential for building a decarbonized and, due to its decentralized structure, more resilient energy system, and are particularly important for Ukraine, where recent conflicts have damaged more than half of its electricity and heat supply capacity. Favorable solar irradiation conditions make Ukraine a strong candidate for large-scale PV deployment, but effective policy requires detailed data on spatial and temporal generation potential. This study fills the data gap by using open-source satellite building footprint data corrected with high-resolution data from eastern Germany. This approach allowed accurate estimates of rooftop area and PV capacity and generation across Ukraine, with simulations revealing a capacity potential of 238.8 GW and a generation potential of 290 TWh/a excluding north-facing. The majority of this potential is located in oblasts (provinces) across the country with large cities such as Donetsk, Dnipro or Kyiv and surroundings. These results, validated against previous studies and available as open data, confirm Ukraine's significant potential for RTPV, supporting both energy resilience and climate goals.

**Keywords**: World Settlement Footprint, Microsoft GlobalML, RESKit, 3D building data, distributed generation

## 1 Introduction

Ukraine's energy system has been under pressure from attacks on key infrastructure since the Russia-Ukraine conflict began in February 2022. According to the International Energy Agency (IEA), Ukraine lost more than half of its electricity and heat generation capacity in the course of 2022-2023, and a further 9 GW in the intensified attacks of spring 2024, leaving it with about a third of its pre-conflict generation capacity [1]. Scientists argue that any effort to rebuild the electricity system should meet four criteria: rapid reconstruction, increased resilience, reduced dependence on fuel imports, and reduction of polluting emissions. All four criteria could be met by a large-scale deployment of high quality solar resources in Ukraine [2], [3], [4], [5] [6]. In its energy action plan for Ukraine, the IEA recommends increasing and decentralizing power supply [1], and highlights solar PV, including behind-the-meter installations, as an option that can be rapidly deployed [1]. Furthermore, decentralization of the energy system is generally considered the best solution to increase resilience [7]. As early as 2022, the World Bank has called for distributed renewable energy generation [8] as a key element of local power system resilience [9]. In addition, the deployment of solar PV as a core element to increase renewable energy production in Ukraine [10] is supported by the national energy policies to move away from fossil fuels [7] and the rapid growth of pre-war solar production, which increased by 114% between 2019 and 2020 [11]. Scientific research supports these recommendations, showing that solar prosumer concepts in Ukraine not only increase energy security [12], but are also economically viable, with profitability increasing with the share of self-consumption [13].

By the beginning of 2024, nearly 1.5 GW of solar photovoltaic (PV) have already been installed by consumers, and in addition to small modular gas turbines, the Ukrainian government has prioritized the deployment of rooftop photovoltaic (RTPV) in non-residential buildings [1]. While decentralization in most other countries is a result of sustainability efforts, decentralization in Ukraine is driven by energy security concerns [1]. Even if solar PV on its own lacks essential system services such as reserve and peaking capacity, it can be combined with for example decentral battery storage or operate as part of a wider array of distributed resources to provide such much-needed services. However, studies suggest an even stronger post-war development of PV [6], suggesting an important role of decentralized PV and in particular RTPV in a future Ukrainian energy system, making it a no-regret option.

When analyzing the RTPV potential, both the spatially resolved capacity potential and the temporarily resolved generation potential are relevant. While the capacity potential provides an estimate of the total installable capacity in a region based on roof availability, the generation potential provides insight into the total electricity that can be generated, taking into account seasonal and daily availability patterns due to solar irradiation fluctuations throughout the year and variations between regions. RTPV capacity potential can be calculated using bottom-up and top-down approaches. Bottom-up approaches are based on roof geometry data, PV module technical parameters, and solar irradiation data. This calculation can be done either for small sets of buildings (e.g., for the Ukrainian cities of Ladyzhyn and Chortkiv [14], [15]) or for larger sets of building types that include the cities themselves [16], [17], [18]. Various simulation tools exist for PV potential calculation [19] and solar modelling [20]. For policy development, however, large-scale RTPV potentials are relevant, i.e., the sum of RTPV potentials of individual buildings in a region. Their calculation requires scalable approaches with input data available at the geographic extent of interest. Some studies simulate the potentials based on detailed 3D building data at level of detail 2 (LoD2), i.e., buildings are represented by blocks with simplified roof shapes [21]. Other studies use deep learning approaches to segment roofs from satellite imagery and simulate RTPV potentials for the detected roofs [19], [22], [23], [24], [25]. In addition to individual cities and small to medium-sized countries [20], these approaches are also applied to very large regions, such as China [26] Europe [27] or the whole world [28] [27]. As an alternative to bottom-up approaches, RTPV capacities can be estimated using top-down approaches. Top-down means that national estimates are scaled based on proxies such as population size or aggregated built-up area, which are then applied on a national scale [29].

There are few studies on RTPV potential in Ukraine (see Table 1). No regional studies at the oblast (Ukrainian province) level could be found and only a single potential assessment with national scope. For the national assessment, Semenuk [30] calculated the available rooftop area of residential buildings using the aggregated footprint data of the State Statistics Service of Ukraine [31], which was drawn from the statistical publication "Housing stock of Ukraine" [32]. In addition, the author assumed that 70% of the rooftops were suitable for the installation of solar panels. Apart from the national study, several local potential and feasibility studies with a higher level of detail could be identified to serve as references for the plausibility check. In particular, the methodology used for the calculation of the city of Zhytomyr [16] included the estimation of rooftop areas using the Public Cadastral Map of Ukraine [33] and the deployment of roof measurements. This required the calculation of the available roof space for apartment blocks, building types, and solar capacity per unit area. For the rooftop potential of the city of Enerhodar [17], the calculation included nearly all buildings with suitable roof structures (both flat and pitched) [17]. An additional set of buildings in the city of Ladyzhyn was studied in detail by physically analyzing each roof structure and modelling potential solar installations [14]. In the European study by Bódis et al. [27], the Ukrainian RTPV potential was estimated via population density as a proxy based on Polish values, assuming similar regional building structures. For the global assessment by Joshi et al. [28], a global fishnet grid was first generated using a top-down approach in ArcGIS PRO. Each cell was then processed on Google Earth Engine to calculate satellite-derived built area, population, and conversion factors. This final bottom-up approach involved the aggregation of individual building footprints, road lengths, and population data for each fishnet cell to provide a more accurate estimate.

**Table 1.** Overview of existing RTPV potential studies for Ukraine.

| Spatial scope | Scope | Methodology | Capacity Potential | Avg. CF | Energy Potential | Ref. |
|---|---|---|---|---|---|---|
| International | Non-EU East (UA, ML, BEL, GE, AZ, AR) | Hybrid | - | - | UA: 294 TWh/a | [28] |
| | EU | Bottom-up | - | - | UA: 33 TWh/a | [27] |
| National | Ukraine (residential buildings) | Top-down | 233.6 GW | 0.15 | UA: 307 TWh/a | [30] |
| Local | Zhytomyr | Bottom-up | 846 MW | - | - | [16] |

| | | | | | |
|---|---|---|---|---|---|
| Sumy (municipal buildings) | Bottom-up | 17.6 MW | - | 20.6 MWh | [18] |
| Enehodar (no industry) | Bottom-up | 86 MW | | | [17] |
| Chortkiv (school no. 5) | Bottom-up | 0.2 MW | - | - | [15] |
| Ladyzhyn (schools + municipal buildings) | Bottom-up | 1.65 MW | - | - | [14] |

Most of the studies listed in Table 1 are limited in spatial scope and resolution. None of the articles provide open access to high spatial resolution capacity potentials or solar generation time-series. However, rapid and cost-effective redesign of a highly decentralized energy system for Ukraine requires detailed and reliable knowledge of spatially resolved capacity limits and temporal patterns of potential energy yields. Replicability and accuracy through a detailed bottom-up approach, including the publication of interim results, using openly available geodata, further contributes to building confidence in the results required for real-world applications [34]. Therefore, the objective of this work is to produce high-resolution RTPV capacity potentials and generation datasets for Ukraine with high reliability, available as open-source data.

This paper is organized as follows: Section 2 describes the bottom-up methodology used to calculate RTPV capacity and generation potential for Ukraine. Section 3 presents the spatially resolved RTPV capacity potential and the spatially and temporally resolved electricity generation potential. Section 4 discusses the presented results in relation to previous studies and the implications for the Ukrainian energy system.

## 2 Methodology

This section describes the methodology used to estimate RTPV capacity and generation potential in Ukraine. First, an overview of the overall workflow is provided (see Section 2.1). Then, the individual steps are described in more detail (see Section 2.2-2.4).

### 2.1 Workflow Overview

In order to provide a high-resolution, highly reliable RTPV potential dataset, a bottom-up approach combined with thorough validation was used. Bottom-up here refers to the aggregation of regional and national potentials in Ukraine based on individual building footprint polygons. The key challenge is the lack of high-resolution ground truth data for Ukraine. To the best knowledge of the authors, neither large-scale official cadastral geodata with building footprint information nor actual roof areas or 3D building data, which are crucial for high resolution and reliability, were available for Ukraine at the time of this study. Therefore, the approach presented in this study uses several factors to translate Ukrainian satellite-based building polygons from two different sources into maximum RTPV capacity potentials per building polygon before aggregating the potentials and simulating the potential energy yield (see Section 2.3 and Figure 1 on the right). These factors can be divided into a footprint-area correction factor per data source and two translation factors, the footprint-to-roof-area translation factor and the roof-area-to-capacity translation factor. These correction and translation factors were calculated using East German ground truth data (Section 2.2) before applying them to Ukrainian satellite data (see Section 2.3). All of these factors depend on the building usage type, i.e. whether the building is residential or non-residential, and the degree of urbanization of the respective raion (Ukrainian district) or city to reflect different settlement structures.

The footprint correction factor was then applied to the area of each satellite-based building polygon in Ukraine to scale it to the actual size of the building footprint, thus correcting the error that may have been introduced by the processing of the satellite imagery. The footprint-to-roof-area translation factor was then multiplied by the corrected footprint area of each polygon to obtain the roof areas. This step is necessary because the ratio of roof area to footprint area depends on the shape of the roof. Finally, the roof-area-to-capacity translation factor provided the maximum installable RTPV capacity for each polygon based on the calculated roof areas. It includes a reduction for roof ineligibilities such as building shape or shading, chimneys, skylights, attics, and other ineligibilities, but also accounts for the installable capacity per eligible roof area considering flat or mounted panel installation. Finally, the resulting polygon capacities were aggregated at district level for each building usage type and as a total capacity. Depending on the degree of urbanization and the respective building usage type, the capacity was assigned to the respective roof slopes and azimuths using statistical distributions of roof slopes and azimuths from East German data. These were grouped into 25 reference configurations per district, which were then used for energy simulations: The potential hourly yield was simulated separately for each system configuration and district using the open-source tool RESKit [19] (see Section 2.4).

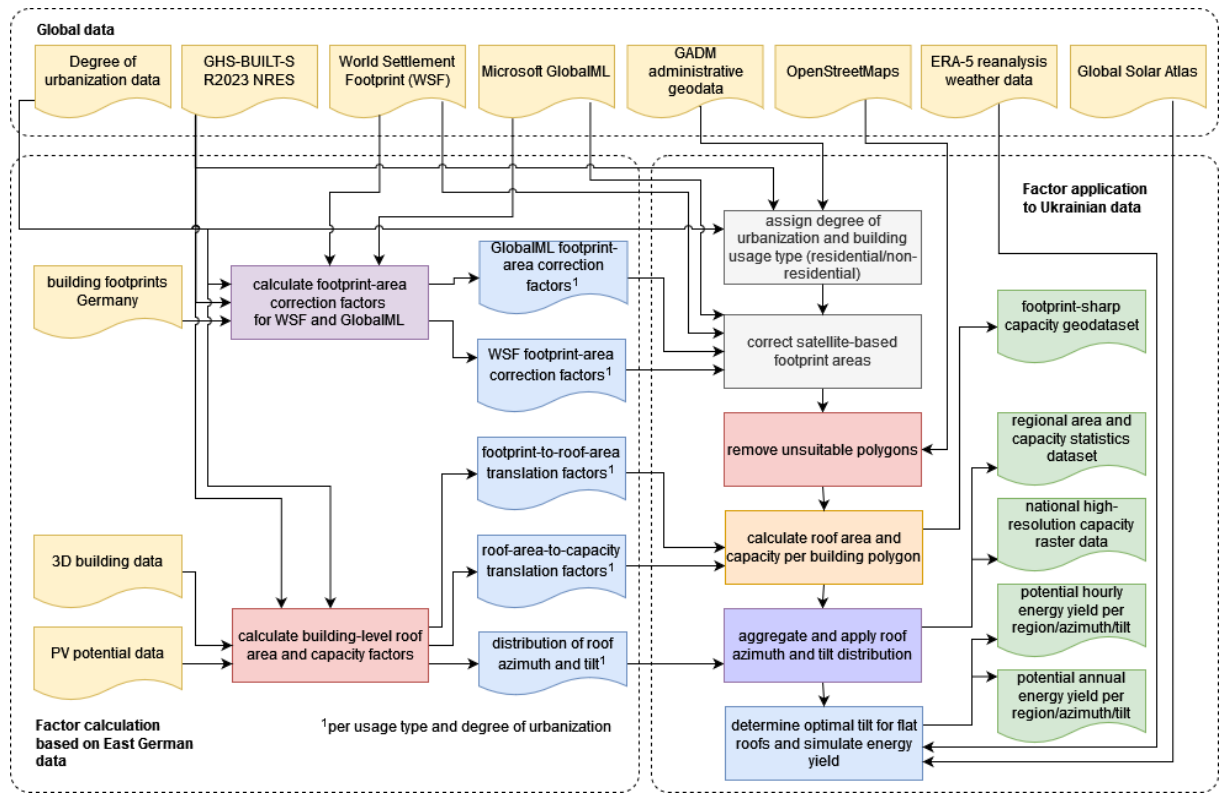

**Figure 1.** Simplified schematic flow chart of the methodology applied in the capacity and energy potentials assessment.

The methodological approach can therefore be divided into two blocks, as visualized in Figure 1: the factor extraction based on German and international training data (left side), and the application of these factors to satellite data of Ukrainian buildings to derive national capacity potentials with subsequent simulation of energy yield (right side). The following sections present in detail the extraction of factors from German data (Section 2.2) and the application of these factors to the Ukrainian context (Section 2.3). Finally, the simulation approach for a temporarily resolved RTPV generation potential is described (Section 2.4). Each of these methodological steps is complemented by a plausibility check developed to support the results of this study (details are provided in Supplementary Material S5). Therefore, real data for Ukraine were collected from national and international literature and supplemented with information from email exchanges with the authority and published reports. The collected real data were then compared with the respective intermediate results such as rooftop area, capacity potentials and energy potentials in the respective subsections of Section 3 at both building and different administrative levels.

## 2.2 Correction and Translation Factors from German High-Resolution Data

As described in Section 2.1, East German building data was used to derive factors that allow estimation of RTPV capacity from available footprint data in Ukraine. In particular, training data from East Germany was used to maximize similarity to Ukrainian building structures, as the 2012 census revealed that over 40% [35] of East German building stock dates back to the former communist German Democratic Republic, which had a strong similarity to the Soviet Union in its building style [36]. East Germany includes the federal states of Brandenburg, Mecklenburg-Western Pomerania, Saxony, Saxony-Anhalt, and Thuringia. An illustration of the correction and translation factors calculated based on East German data is shown in Figure 2. To correct the Microsoft GlobalML [37] and World Settlement Footprint (WSF) [38] datasets available for Ukraine, two separate footprint-area correction factors were derived, hereafter referred to as correction factors $c_{globalml}$ and $c_{wsf}$, respectively. In addition, footprint-area-to-roof-area and roof-area-to-capacity translation factors were calculated, hereafter referred to as translation factors $t_r$ and $t_{pv}$. The methodology for extracting the factors is described in detail in Supplementary Material S1.

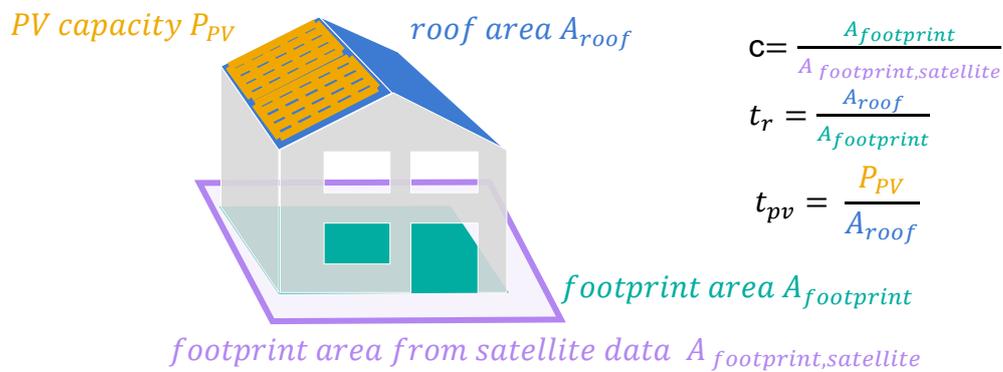

**Figure 2.** Illustration of footprint-area correction and footprint-to-roof-area and roof-area-to-capacity translation factors calculated from East German datasets.

## 2.3 Determining RTPV Capacity in Ukraine

After extracting the footprint-area correction factors as well as the footprint-to-roof-area and roof-area-to-capacity translation factors based on East German training data, these correction and translation factors were applied to Ukrainian footprint data based on high-resolution satellite imagery. Analogous to the above extraction of correction and translation factors, the capacity assessment for Ukraine was also processed at the corresponding Ukrainian administrative unit of "raion" or city level, both of which will be referred to as "districts" in the following for readability. The Ukrainian region shapes were therefore extracted from GADM.org [39] at the respective second administrative level (GID-2) and then intersected with the building footprint vector data. Only footprint polygons with a centroid within the district geometry were considered. GlobalML [35] was used as the primary data source for building footprints whenever its coverage exceeded 99% of the district area, which was the case for 77% of all regions (see Supplementary Material S4 for a district-level map). For the remaining regions, the World Settlement Footprint (WSF) [38] was used after polygonization. To apply the correction and translation factors, the respective degree of urbanization was then assigned to each district by matching the "GID_2" codes with the urbanization data defined by the Global Degree of Urbanisation Classification of administrative units (R2023) from the Global Human Settlement Layer (GHS-DUC) [36], introduced in Section 2.2. Next, the footprint polygons were each assigned a building type (residential/non-residential) based on the global raster dataset GHSL-SBUILT-S-NRES [40], analogous to the processing of the German training footprint data (see Section 2.2). Based on the footprint vector data source, the building usage type, and the degree of urbanization, the respective footprint-area correction factors were applied to each building polygon area in order to obtain realistic footprint areas. In a subsequent step, ineligible building structures were filtered out by geospatially matching them with ineligible building type polygons from OpenStreetMaps [37]. The data were downloaded from the

Geofabrik download server [38]. Such inappropriate building types include dams or towers, but also historic castles and abandoned buildings, or small features such as bus stops. A list of all building and land cover types that were deemed irrelevant for the analysis is available in Supplementary Material S2. Based on the individual footprint area and building usage type, as well as the level of urbanization of the district, the respective footprint-to-roof-area and roof-area-to-capacity translation factors were then applied to the corrected building footprint areas to derive the roof area and maximum installable RTPV capacity per each building polygon. The individual building capacity potentials were then aggregated at the district level, separately for residential and non-residential structures. Finally, the total district capacity per usage type was distributed across the different azimuthal orientations and roof tilts according to the representative relative distribution derived from East German building data (see Supplementary Material S1 and S3) for the respective degree of urbanization and building usage type.

## 2.4 Simulating RTPV Generation Potential of Ukraine

The maximum solar energy yield of RTPV per district, roof slope, and azimuth was simulated using the open-source tool RESKit [19]. At the centroid of each district shape, 25 representative synthetic solar RTPV installations were defined, each representing a combination of roof tilt and solar azimuth. For each of the 10° roof tilt angle bins (see Table 1 in Supplementary Material S1), the corresponding average tilt value was used for the simulation. The capacity of each synthetic plant is equal to the district capacity of the corresponding roof tilt and azimuth bin. Roofs with a tilt of less than 10° were considered as flat roofs. For such flat roofs, a combination of south-facing orientation and location-optimal tilt angle was assumed, the latter defined by the RESKit.location_to_tilt() method following Ryberg et al [19]. Each of these 15,725 representative solar plants was then simulated in RESKit using ERA-5 weather reanalysis data [43] for hourly resolution and long-term average insolation from the Global Solar Atlas [44] for spatial disaggregation to 250 m x 250 m cells. Capacity factors were simulated for 20 years from 2000 to 2019 and then averaged per installation to avoid bias from extreme weather years. The average capacity factors were then used to estimate an average levelized cost of electricity (LCOE) for the year 2030. For this purpose, a CAPEX of 705 EUR/kW$_p$ and an OPEX of 9.80 EUR/(kW$_p$*a) were assumed as an average of the residential and non-residential RTPV cost estimates provided by the Danish Energy Agency for the year 2030 [39]. The economic lifetime was assumed to be 20 years and the weighted average cost of capital (WACC) was assumed to be 8% per year. The technical parameters of the modules were interpolated for the year 2030 based on the 2050 projection of the WINAICO WSx-240P6 PV module defined by Ryberg [46].

# 3 Results

This section presents the results of applying the factors calculated on the basis of East German data (presented in Supplementary Material S3) to Ukrainian data. It includes a detailed assessment of the resulting high-resolution RTPV capacity data (Section 3.1) as well as of the electricity generation potential (Section 3.2). Furthermore, the supplementary material provides a plausibility check for all the above-mentioned aspects, including the footprint areas of Ukrainian cities (see Supplementary Material S5).

## 3.1 Capacity potential

Ukraine's RTPV capacity potential was calculated using a bottom-up approach at the individual building level (see Figure 3b) and was found to be 238.8 GW, excluding north-facing capacity. The north-facing capacity in Ukraine would add another 48.2 GW, but is not considered here as part of the economic potential. The national capacity pattern is largely consistent with population density, with the potential in large cities highlighted in Figure 3a.

.

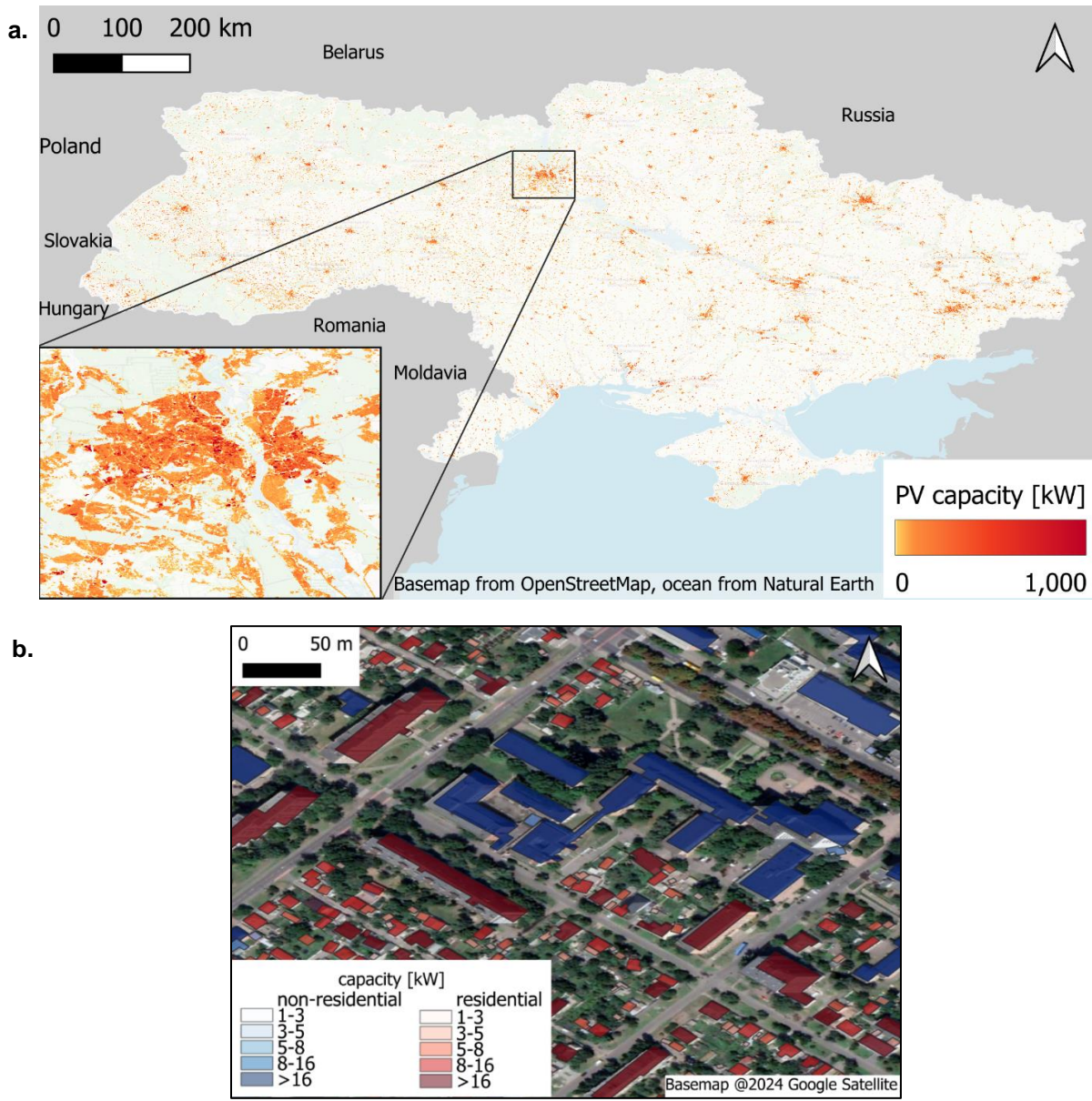

**Figure 3**. a. National RTPV capacity distribution raster, including North orientation. Map extract in the bottom left corner shows zoomed-in data of Kyiv. b. Example extract for high-resolution capacity results based on GlobalML data for residential (red) and non-residential (blue) buildings in Cherkasy.

The potential is therefore unevenly distributed across the 27 oblasts, with only 2 oblasts (Kyiv incl. surroundings and Donetsk) exceeding 20 GW regionally, driven by their respective dominating cities and size. The oblasts with the largest potential also host the largest cities, namely Donetsk, Dnipro, Kyiv, Odesa, Kharkiv and Lviv. Sevastopol ranks last with 1.56 GW due to its small size, but the pattern changes significantly when capacity is related to oblast area: Especially Kyiv City stands out with an exceptionally high potential area capacity density of 6,351 MW/km², but Sevastopol comes in second with 2,004 MW/km², whereas the remaining oblasts range between 261 and 897 MW/km². Detailed results per oblast are presented in Supplementary Material S4. The differences are even more extreme at the district level, where values exceeding 16,700 MW/km² are reached in central Lviv or Odesa, while rural districts (raions) may have less than 100 MW/km². The district median is 388 MW/km² and almost ¾ of all districts are below 1,000 MW/km². Local differences are also evident when comparing the distribution of capacity density by degree of urbanization. Urban districts contribute more than 21% of the national RTPV potential, although they represent only 15% of the districts. The median capacity density of urban districts, at 5,600 MW/km², is more than 20 times higher than the median capacity density of rural districts, at 288 MW/km², and still almost 14 times higher than the median capacity density of suburban districts, at 403 MW/km². However, the latter districts are characterized by the largest spread due to several outliers up to 8,262 MW/km², resulting in an average capacity density of 1,139 MW/km², which is significantly higher than the median. A plot of the distributions by degree of urbanization and a district-level capacity potential map are provided in Supplementary Material S4.

The module orientation follows the statistical distribution of the derived roof angles as described in Supplementary Material S1. It should be noted, however, that the optimal design for flat roofs is south-facing, which increases the south-facing portion to approximately 52% of the total capacity, with the remainder almost evenly distributed between north, west, and east orientations.

The methodology for assessing the plausibility of Ukraine's RTPV potential was developed to support the results of this study. To ensure the accuracy of the assumptions, data was collected from literature for Ukraine, considering various types of buildings such as residential one-family houses, residential flats, administrative buildings, and industrial buildings. Semenuk [30] estimates the potential for rooftop solar panels in Ukraine to be 233.6 GW, based on the country's residential building stock. In the case of individual houses, it is assumed that a RTPV system with a capacity of 25.6 kW can be installed on each building based on findings of the Ukrainian State Agency for Energy Efficiency and Energy Saving [16], [30]. In addition, data for apartment blocks have been included. The slightly higher total Ukrainian RTPV capacity of 239 GW presented in our article aligns very well with Semenuk's findings, although Semenuk excludes municipal and industrial buildings. The potential of RTPV for industry and service sector for Zhytomyr is approximately 37.5 MW [16]. Accordingly, the potential of RTPV for apartment buildings is approximately 135 MW. The authors also assumed that at least half of the private houses in the city can accommodate a 25.6 kW RTPV, which is the current national average in Ukraine. Thus, the theoretically possible potential of rooftop solar installations for private houses is 336 MW. Therefore the total RTPV potential of the Zhytomyr city is around 508 MW [16]. Our calculation suggests a potential RTPV capacity of around 846 MW for the city. The difference can be explained in wide parts by the lower capacity per rooftop area assumption of only 50 W/m² combined with an underestimation of the industrial rooftop areas due to a manual assessment in the Zhytomyr study [16]. The total RTPV potential of the city of Enerhodar is about 86 MW [17], of which 21.6 MW comes from municipal buildings. Our analysis shows that the RTPV potential of the city is about 74 MW with a roof area of 292,980 m². These slight differences might be explained by additional suburbs which were not considered part of Enerhodar city in the present assessment. Not only have entire cities been estimated in terms of their rooftop/built-up area and RTPV potential, but also specific groups of buildings for selected cities. According to the "Charitable Foundation Evaluating Solar Potential for Municipal Facilities in Ladyzhyn City," there are 18 municipal buildings situated within the city limits, including educational institutions such as schools and gymnasiums, that were assessed to ascertain the optimal and maximum rooftop potential. The aggregate potential of these buildings is estimated to be 1.65 MW [14]. Given that the assessment made use only of larger flat and South-facing roof sections, the value aligns well with the findings of this study which yield 3.9 MW for the whole usable roof area including all azimuths for the aforementioned 18 buildings.

## 3.2 Energy potential

The following subsections present the results of the energy yield simulations, starting with the achievable full-load hours before moving on to the simulated electricity time-series. Then, cost and total energy potentials are discussed before closing with a plausibility check subsection.

### 3.2.1 Annual Full-Load Hours

The energy yield of RTPV in Ukraine varies considerably across the country (see Figure 4). It shows the highest yields for sites in Sevastopol on the Crimean Peninsula, reaching an average of 1,583 FLH/a, while sites in the mountainous climate of the Carpathian Mountains in western Ukraine can only reach about 1,178 FLH/a. With the exception of these mountainous areas, there is a general increasing trend in full load hours from the northwest to the coastal plains in the southeast.

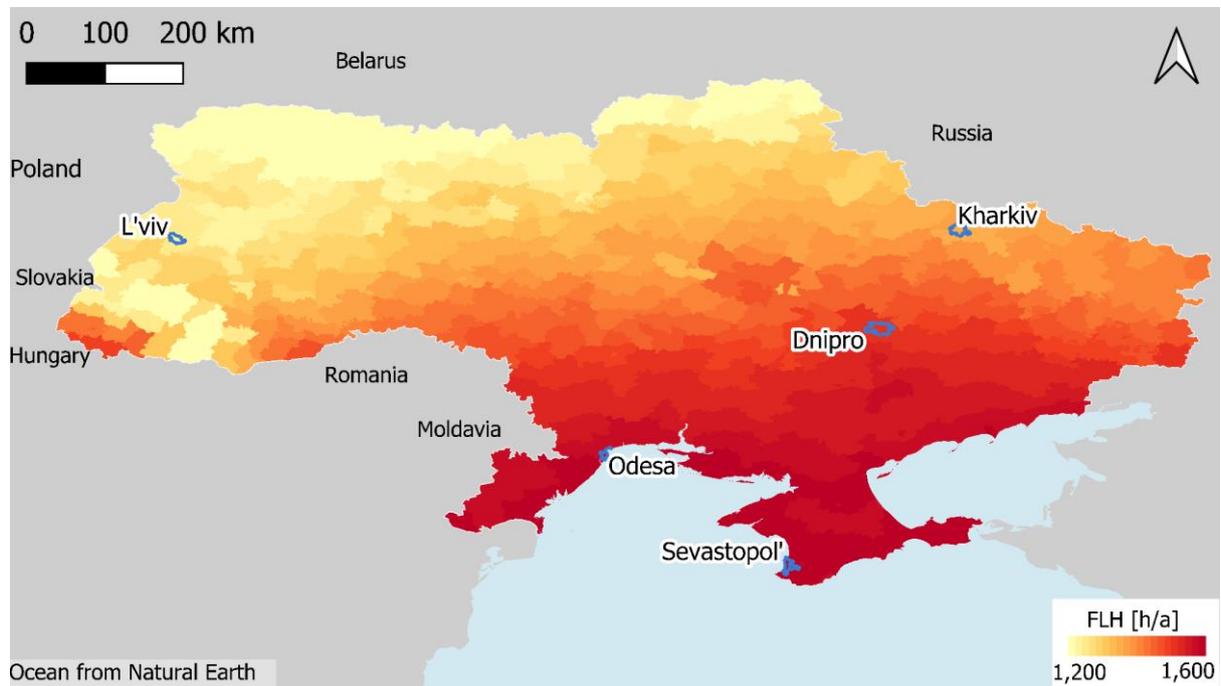

**Figure 4.** National map of full load hours per district for a RTPV system with optimal tilt and azimuth.

Figure 5 ranks all 629 Ukrainian districts by their RTPV yield, highlighting outliers at the low end (Carpathian districts) and high end (Western Crimea). The majority of all Ukrainian districts fall between 1,239 and 1,457 FLH/a (5th-95th percentile) for optimal module locations. In addition to regional differences, the plot also shows large differences between different azimuthal and tilt orientations. South-facing plants with a tilt angle of 30-40° are close to the yield-optimal module tilt angle, which was found to be between 36.5-39.8° across Ukraine. The yield decreases with decreasing module tilt angle for south-facing systems, at 15° tilt angle about 6-7% less energy can be expected compared to the optimal tilt angle. However, even the lowest-yielding south-facing systems outperform other azimuthal orientations by an average of 136 FLH/a. East and west facing systems perform similarly, with a slight advantage for east facing modules. However, unlike south-facing systems, the highest yields are associated with the lowest tilt angles for east- and west-facing systems, as well as for north-facing systems. Here, energy production decreases with increasing tilt angle, which is most pronounced for north-facing systems, which lose about half of their yield at 50° tilt compared to 10°.

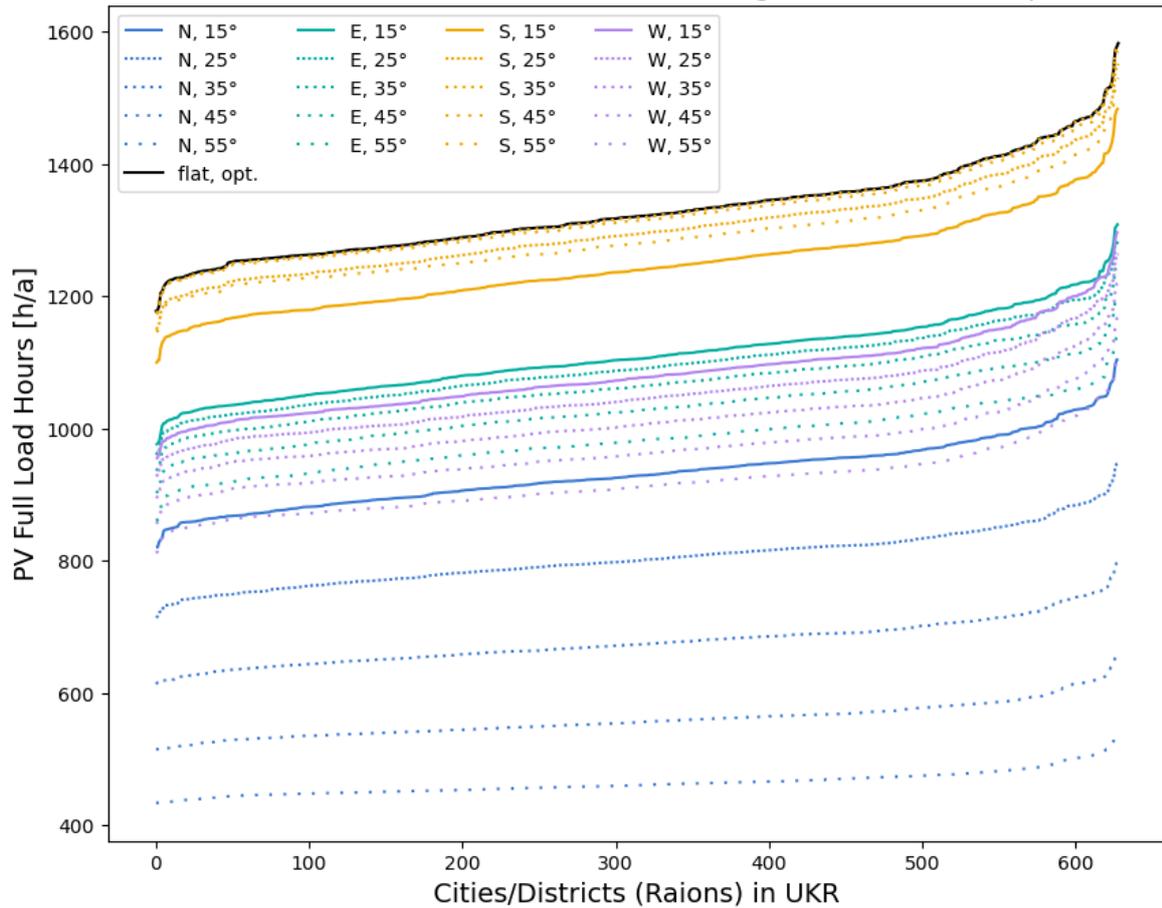

**Figure 5.** Achievable full-load hours for every district in Ukraine for different azimuthal and tilt configurations.

To show inter-annual variations, five representative cities were selected to visualize patterns in different regions of the country: Lviv in the west, Kharkiv in the northeast, Dnipro in the center, Sevastopol on the Crimea, and Odesa in the south. Very similar trends can be observed in the central belt from Odesa via Dnipro to Kharkiv, while Lviv in the far west and Sevastopol on the Crimea deviate slightly from this pattern in some years (see Figure 6). The annual averages for the shown optimal system configurations deviate the least in the coastal regions with a variation of about 7% against 9-14% in the other sample regions. On a national scale, the annual variations between 2000 and 2019 are particularly high in the northeast of the country and to a lesser extent near the Carpathians (see Supplementary Material S4). For example, 2007 was a very good solar year everywhere except in the west of the country, while 2018 was excellent in most parts of the country, but average along the coast in the south. Comparing the annual averages of all 15,725 systems in Ukraine with their respective 20-year average FLH value, 2002 emerges as the year that is closest to the long-term average FLH and thus representative of most systems, and also shows minimal deviation from the respective long-term average FLH in all systems.

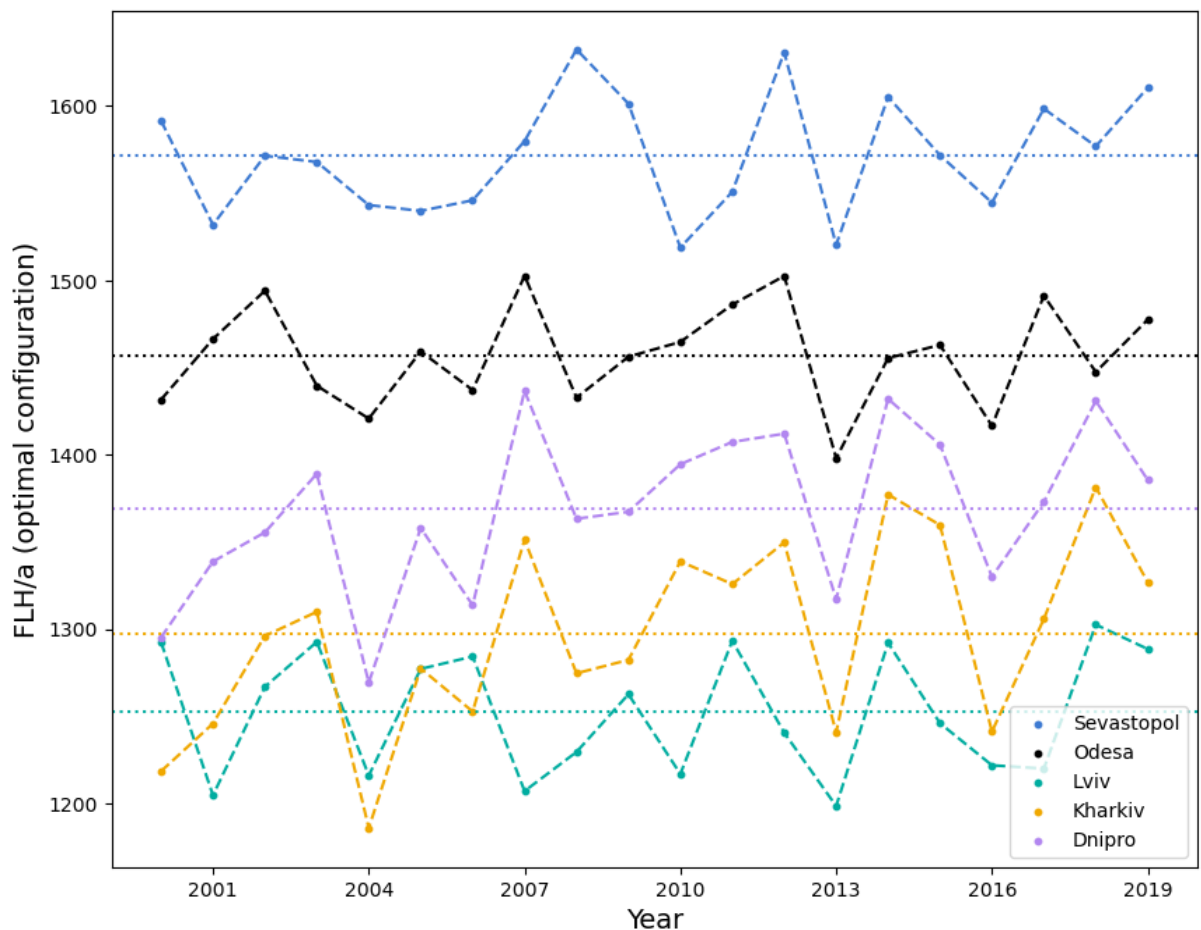

**Figure 6.** Annual FLH for representative locations and an optimal system configuration over a span of 20 years.

The 20-year national average FLH, when all possible district and system configurations are weighted by their energy contribution, is 1,237 FLH/a (1,184 FLH/a if north orientation is also considered). 1,332 FLH/a can be achieved on average by south-facing systems only.

To ensure the accuracy of the solar energy potential through plausibility checks, national and international publications are reviewed, including data from similar countries in Eastern Europe. In cases where direct data for Ukraine are not available, the energy potential is approximated by using data from Eastern European countries and adjusting the figures based on the population of Ukraine (see Supplementary Material S5 for more details).

### 3.2.2 Electricity Production Time-series

Apart from the average annual power output, the system configurations also differ considerably in their respective power supply time series. Figure 7 shows exemplarily the capacity factor time series of systems with different azimuths and tilt angles in Kyiv. The dash-dotted lines indicate an average capacity factor per azimuthal orientation, always for the tilt angle with the higher yield. The upper part of the figure shows the hourly time series of a typical summer (a) and winter (b) day with very little cloud cover. The south-facing system has a symmetrical peak around noon, while the west- and east-facing systems peak in the morning and afternoon, when south-facing production is still low. While tilt angle has a limited effect on south-facing systems in the summer, the effect is much greater on east-west systems, where a 55° tilt angle both increases yield and shifts the peak to the morning/evening. North-facing systems with high tilt angles, on the other hand, have extremely low yields and benefit mainly from the morning and evening sun, with lower yields around midday. North-facing systems with low tilt angles behave similarly to south-facing systems, but with a lower overall yield. The average daily yield of the east-facing system even slightly exceeds that of the south-facing system in summer, while the west-facing system suffers from possible cloudiness around 4 pm. Except for the steep north-facing modules, there is no big difference between the average summer capacity factors of the different systems. In winter, however, the south-facing systems excel by far, with higher yields at higher inclinations, albeit at less than half the average summer capacity factor. The sun rises too late and sets too early for east- and west-facing systems to take advantage of its morning and evening peaks. This is also reflected in the lower part c) of Figure 7, which shows the average weekly capacity factors over the year. The expected seasonal trend of strong summer production and weak winter production is clearly visible for all systems. While summer yields are fairly similar for all systems except steep north-facing modules, south-facing systems perform considerably better in the low season between August and late April. Flat east-/west-facing systems outperform their steeper tilt counterparts. The annual average capacity factor of south-facing systems, especially those with a steep tilt angle, is still about 25% higher than the best east-/west-facing modules.

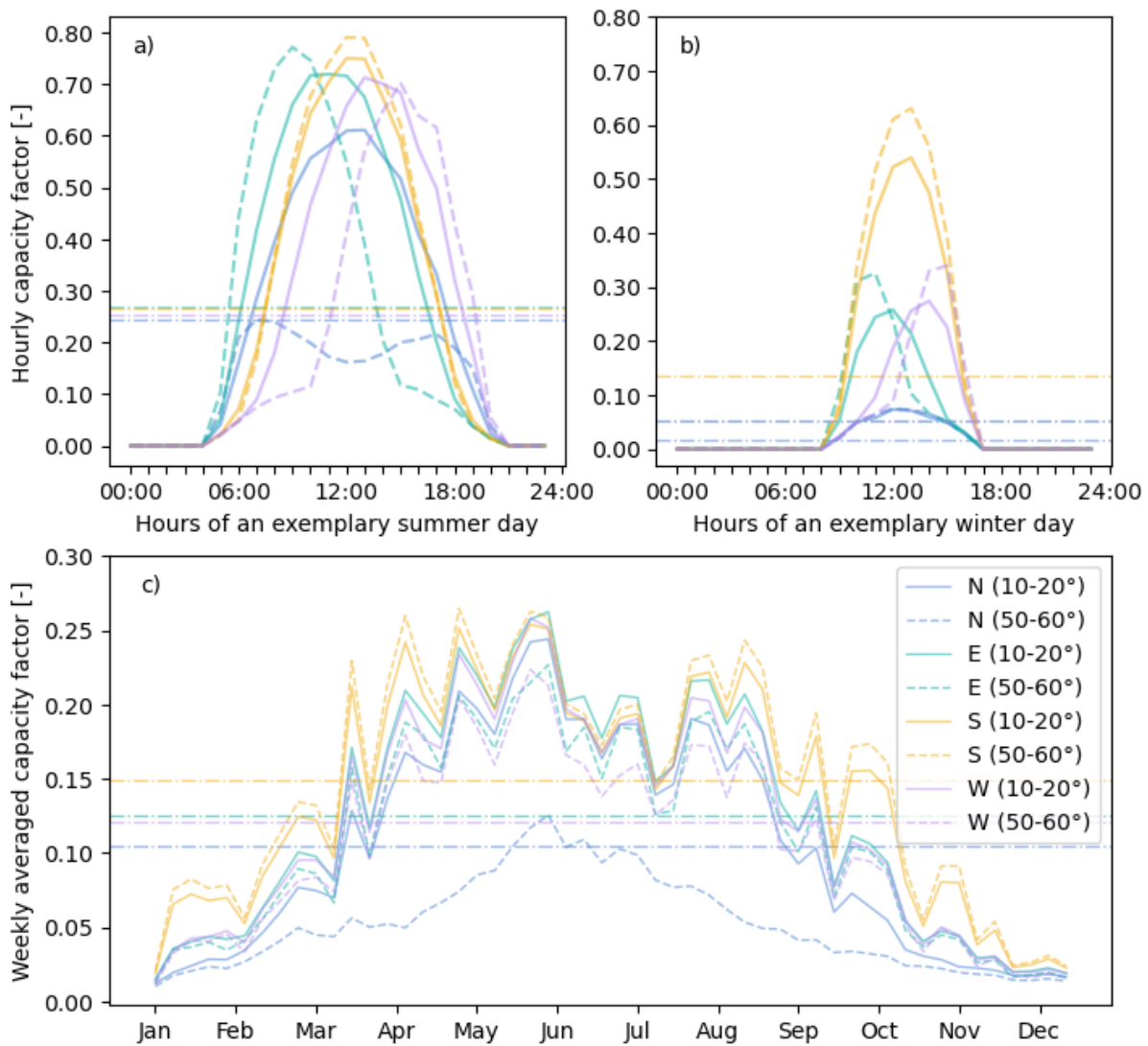

**Figure 7.** Exemplary hourly and annual timeseries for different RTPV system configurations in Ukraine.

### 3.2.3 Cost and Energy Potential

Applying the simulated full load hours to the above capacity potentials results in a total energy potential of 290.1 TWh for RTPV in Ukraine, again excluding north-facing systems, which would add another 32.6 TWh. 68% of the electricity could be generated by south-facing systems, including flat roof potential, with the remaining 32% split almost equally between east- and west-facing systems. More detailed values per oblast can be found in Supplementary Material S4.

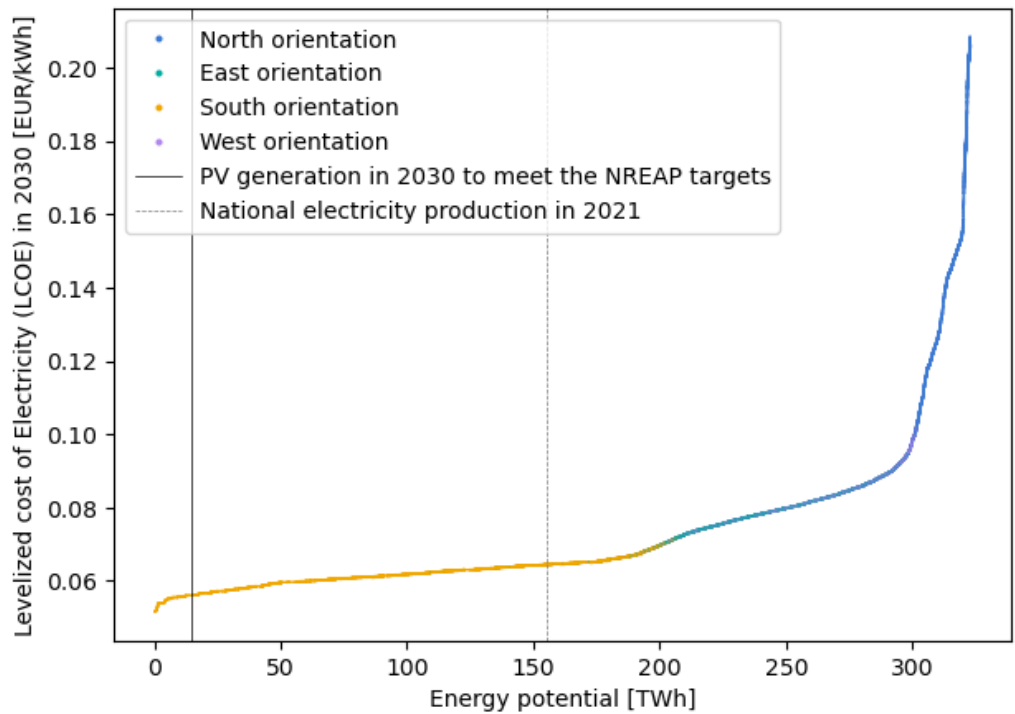

**Figure 8.** Cost-potential curve for RTPV incl. North orientation in Ukraine in 2030, averaged over 20 weather years.

93% of the total potential, including north-facing sites, could be realized below 10 Ct€/kWh and 78% below 8.0 Ct€/kWh by 2030, as shown in the cost-potential curve in Figure 8. These costs should be understood as the marginal levelized cost of electricity (LCOE), without any value adjustment to account for the limitations of solar PV technology, such as volatility and lack of reserve or peaking capacity. The lowest cost potentials are achieved with south-facing systems, starting at around 5.2 Ct€/kWh. Apart from a few initial outliers, the cost increase with increasing potential expansion is nearly linear at the national level, adding about 0.5 Ct€/kWh per 100 TWh of potential expansion, averaged between the first and third quartiles. 61% of the national energy potential, well above the pre-war electricity production in 2021, is contributed by south-facing systems (including flat roofs). A small step in the cost function around 190 TWh indicates the addition of more expensive east- and west-facing systems to the mix; costs also increase faster for east-west facing systems, averaging 2 Ct€ per 100 TWh/a of potential expansion. This is followed by the cheapest energy supply from north-facing systems with very low tilt angles at 7.4 Ct€/kWh. Northern azimuths start to dominate the additions at around 8.1 Ct€/kWh, causing a massive increase in marginal costs to above ca. 9 Ct€/kWh to over 20 Ct€/kWh without significant further energy contribution.

### 3.2.4 Energy and cost potential validation

The LCOE calculator developed for Ukraine by the Clean Energy Lab [48] shows a range of 6.4-8.5 Ct€/kWh for the provincial average LCOE under the same CAPEX and discount rate assumptions, which is slightly higher than the costs determined in this study for 2030. The PVGIS tool [49] of the Joint Research Center of the European Commission yields on average about 12% less electricity production/higher costs for the optimal module orientation, with little variation between districts. The deviation can be explained by the improved module efficiency, which was projected to 2030. The comparisons thus confirm the magnitude of the above yield and levelized cost results. According to Joshi et al [28], the "non-EU East" countries with a population of 71 million people have a total energy potential of 499 TWh/a. Scaling this to the Ukrainian population of 42.3 million, the energy potential of

Ukraine can be estimated to be 294 TWh/a, which is within 10% of the results of the present work with a total value of 322.7 TWh/a.

## 4 Discussion and Conclusion

This study shows that using openly available spatial building footprint datasets and correcting them to better match actual building footprint areas is a useful approach for regions where ground truth data is lacking. Microsoft GlobalML is shown to be closer to the official building dataset available for Germany in terms of building footprint areas. WSF, on the other hand, provides much larger areas due to its raster format, which makes it unsuitable to capture exact building footprints. Therefore, it is recommended to use Microsoft GlobalML whenever possible. However, the coverage of Microsoft GlobalML must be carefully evaluated, as it occasionally has gaps. The choice of open datasets for the analysis contributes to their reproducibility and transferability. The workflow can be easily adapted to other regions of the world.

The RTPV capacity potential of Ukraine is 238.8 GW (excluding north-facing roofs), which is slightly more than 4 times the total installed electricity generation capacity in Ukraine before the war [1]. Due to its dependence on building surfaces, the RTPV potential aligns well with the population distribution pattern and consequently with the demand centers in Ukraine. Its natural spatial proximity to potential users, be they industrial or residential tenants of the respective buildings, makes it an ideal opportunity for prosumers, thus aligning very well with the goal of a decentralized, resilient energy system [1]. The total energetic RTPV potential of Ukraine, excluding north-facing systems, is about 290 TWh/a, which is almost double the national pre-war electricity production of 155.5 TWh in 2021 [50]. A clear regional trend in full-load hours extends from the northwest to the southeast of the country, with solar yields in the northwest and the Carpathian Mountains about 25% lower than in the districts with the best potential near the Black Sea coast. However, Ukraine's solar potential is very attractive overall, the FLH range is roughly comparable to that of Italy, excluding Sicily and the Alps [51], and even the lowest performing regions are still just below the level of the best performing regions in Germany [52]. South-facing systems account for over 60% of the total energy potential and are available at LCOEs between 5.2-7.5 Ct€/kWh in 2030. The LCOE rises sharply above 9 Ct€/kWh, which includes about 91% of the total potential, including north-facing. 171 TWh/a and thus slightly more than the equivalent of the pre-war electricity consumption in 2021 [50] can be produced below 6.5 Ct€/kWh. A direct cost comparison with other, especially non-volatile, power sources would be inappropriate, as the above LCOE are based on energy only and have not been value-adjusted for aspects such as volatility, seasonality, or lack of reserve and peaking capacity. Nevertheless, a range of typical other electricity generation costs in Ukraine are provided here for context. Without value adjustment, the LCOE of solar power in Ukraine would be significantly cheaper than electricity from coal (12.4 Ct€/kWh), nuclear power (10.5 Ct€/kWh) or open cycle gas turbines (17.5 Ct€/kWh) in 2021 according to the default values of the LCOE calculator tool of the Ukrainian Clean Energy Lab (CEL) [48], [53]. Combined cycle gas turbines have similar LCOE [48], [53] as RTPV, but cannot be directly compared due to their heat component. Gas prices in Ukraine have continued to rise steadily over the past two decades [54] and are unlikely to return to pre-war levels in the near future. However, as mentioned above, the limited hourly and seasonal availability of solar power must be taken into account, especially as electricity demand in Ukraine is much higher in winter [1]. Due to the strong dependence on the demand profiles and the available reserve capacity of the electricity grid, the actual value of RTPV for the Ukrainian energy system should be addressed in a separate system analysis approach.

The analysis of the hourly time series shows a strong seasonality with high summer and low winter production for all regions and system configurations. Significant differences between module orientations can be observed, with an advantage of east- and west-facing in summer, both due to a higher potential yield during the long summer days, when the systems benefit from the extreme solar azimuths at sunrise and sunset, but also because their peaks occur comparatively early in the morning and late in the evening. An east-west orientation, rather than the typical south-facing orientation, could therefore allow for higher levels of self-consumption and thus greater energy security and reduced grid load, as electricity demand rises early in the morning and, depending on the season, even peaks in the evening hours before falling sharply at night [55]. However, this benefit must be balanced against lower winter (and therefore annual) production when the sun rises too late and sets too early, making the angle of incidence too steep for the east- and west-facing module planes. The positive and negative effects are more pronounced at high tilt angles. Depending on battery costs, south-facing systems combined with battery storage may still be the better choice in terms of seasonality, but a winter trough has to be taken into account in any case. North-facing systems have no advantage over other module orientations and should be installed at very low tilt angles if the capacity is required. If possible, other module orientations should be preferred.

The plausibility check shows that the deviations for the rooftop and footprint areas of the Ukrainian cities and individual buildings do not exceed 11% for all cities, with a single outlier of 16% for the city of Poltava. The comparison of the capacity potential shows a very good agreement with most of the studies at the national and city level, only the city of Zhytomyr showed noteworthy deviations, which can be explained by different assumptions and approaches. Also, realistic full load hours, energy yield and levelized cost could be demonstrated by comparing the energy simulation results with different tool outputs and studies. Overall, these results confirm the accuracy of the chosen methodology at each step, both at the global level and at the city level. It should be mentioned that the feasible potential will be considerably lower than the technical potential shown in this work due to aspects such as, for example, investment constraints of individual households or local demand and grid conditions. It is also imperative to mention that due to the ongoing war in Ukraine, significant portions of the derived potential cannot be developed at the time of writing, either due to occupation, military operations, or large-scale war damage to buildings. These effects cannot be scaled due to the rapidly evolving situation. In the meantime, however, the very high resolution of the data provided makes it possible to assess each region or district separately and to build a resilient power supply system that is less susceptible to centralized attacks.

The results of this first national-scale study in the literature, including capacities, energies, and time series, are made available as open data at high spatial and temporal resolution. The openly available raster dataset with high resolution capacities allows flexible extraction of capacities at any customized level of regionalization. This serves as valuable information for spatially targeted policy schemes and can be used as input for further analysis, supporting the rebuilding of a resilient energy system for Ukraine. Due to the high temporal resolution of the time series provided for each district and system configuration, it is possible to model temporal system aspects and to evaluate different roof orientations separately. This can be used, for example, to determine the individual system value and possibly subsidies or remuneration schemes. In summary, this study and the datasets published with it go far beyond what was available for Ukraine in terms of both spatial and temporal resolution to support energy system modeling and planning, and ultimately to restore a resilient energy system in Ukraine.

# Data Availability

All result data is available for download at: https://www.iea.org/reports/empowering-ukraine-through-a-decentralised-electricity-system#downloads. Furthermore, a GUI visualizing the data can be accessed via: https://www.iea.org/reports/empowering-ukraine-through-a-decentralised-electricity-system.

# Declaration of generative AI and AI-assisted technologies in the writing process

During the preparation of this work the authors used DeepL in order to check grammar and spelling and to improve wording. After using this tool/service, the authors reviewed and edited the content as needed and take full responsibility for the content of the published article.

## Acknowledgements

This work was supported by the Helmholtz Association under the program "Energy System Design".

Open Access Publications funded by the Deutsche Forschungsgemeinschaft (DFG, German Research Foundation) – 491111487.

DryHy (BMBF, Förderkennzeichen 03SF0716): The authors acknowledge funding provided by the Federal Ministry of Education and Research (BMBF) within "Project DryHy: water-positive generation of hydrogen and e-fuels in arid regions (phase 1)" (FKZ: 03SF0716).

## Bibliography

[1] "Ukraine's Energy Security and the Coming Winter," International Energy Agency (IEA), Paris, Sep. 2024.
[2] "Photovolotaics Report," Fraunhofer Institute forSolar Energy Systems (ISE), PSE Projects GmbH, Jul. 2024.
[3] I. Doronina et al., "Why renewables should be at the center of rebuilding the Ukrainian electricity system," *Joule*, vol. 8, no. 10, pp. 2715–2720, Oct. 2024, doi: 10.1016/j.joule.2024.08.014.
[4] "Renewable Capacity Statistics 2024," International Renewable Energy Agency (IRENA), Abu Dhabi, 2024.
[5] "Snapshot of Global PV Markets - 2023," 2023.
[6] A. Iwaszczuk, I. Zapukhliak, N. Iwaszczuk, O. Dzoba, O. Romashko, and N. Krykhivska, "Prospects for the Development of Photovoltaics in Ukraine," *ERSJ*, vol. XXVI, no. Issue 4, pp. 308–338, Oct. 2023, doi: 10.35808/ersj/3287.
[7] S. Petrovic, Ed., *World Energy Handbook*. Cham: Springer International Publishing, 2023. doi: 10.1007/978-3-031-31625-8.
[8] "Ukraine: Rapid Damage and Needs Assessment," The World Bank, Aug. 2022.
[9] "Proposals for a green recovery in Ukraine," Green Deal Ukraïne (GDU), Berlin/Warsaw/Kyiv, Dec. 2024.
[10] I. Fedorenko, G. Chornous, I. Didenko, L. Anisimova, and S. Mohyl, "Optimization of Renewable Energy Development Strategies in Ukraine," in *2024 14th International Conference on Advanced Computer Information Technologies (ACIT)*, Ceske Budejovice, Czech Republic: IEEE, Sep. 2024, pp. 264–269. doi: 10.1109/ACIT62333.2024.10712603.
[11] "BP Statistical Review of World Energy," BP p.l.c., 2021.
[12] M. Li, U. Pysmenna, S. Petrovets, I. Sotnyk, and T. Kurbatova, "Managing the development of decentralized energy systems with photovoltaic and biogas household prosumers," *Energy Reports*, vol. 12, pp. 4466–4474, Dec. 2024, doi: 10.1016/j.egyr.2024.10.011.
[13] I. Sotnyk, T. Kurbatova, A. Blumberga, O. Kubatko, and O. Prokopenko, "Solar business prosumers in Ukraine: Should we wait for them to appear?," *Energy Policy*, vol. 178, p. 113585, Jul. 2023, doi: 10.1016/j.enpol.2023.113585.
[14] Repower Ukraine., "Assessment of the Solar Potential of Roofs of Municipal Buildings in the City of Ladyzhyn."
[15] Н. Хоодова, "Analysis of the Renewable Energy Potential of the City of Chortkiv and Chortkiv District".
[16] Zhytomyr City Council, "Information-Analytical Note 'Model Scenario Assessments of Zhytomyr's Transition to 100% Renewable Energy Sources by 2050.'"
[17] Enerhodar City Council, "Sustainable Energy and Climate Action Plan of the City of Enerhodar until 2030.," *European Covenant of Mayors Community*.
[18] V. Lesyuk, "PLAN OF ACTION FOR SUSTAINABLE ENERGY DEVELOPMENT AND CLIMATE FOR THE SUMY CITY TERRITORIAL COMMUNITY".
[19] D. S. Ryberg, H. Heinrichs, M. Robinius, and D. Stolten, *RESKit - Renewable Energy Simulation toolkit for Python*. (Mar. 24, 2022). Python. FZJ-IEK3. Accessed: Apr. 06, 2022. [Online]. Available: https://github.com/FZJ-IEK3-VSA/RESKit
[20] R. McKenna et al., "Exploring trade-offs between landscape impact, land use and resource quality for onshore variable renewable energy: an application to Great Britain," *Energy*, vol. 250, p. 123754, Jul. 2022, doi: 10.1016/j.energy.2022.123754.

[21] S. Risch et al., "Potentials of Renewable Energy Sources in Germany and the Influence of Land Use Datasets," *Energies*, vol. 15, no. 15, p. 5536, Jan. 2022, doi: 10.3390/en15155536.
[22] R. Pueblas, P. Kuckertz, J. M. Weinand, L. Kotzur, and D. Stolten, "ETHOS.PASSION: An open-source workflow for rooftop photovoltaic potential assessments from satellite imagery," *Solar Energy*, vol. 265, p. 112094, Nov. 2023, doi: 10.1016/j.solener.2023.112094.
[23] P. Li et al., "Understanding rooftop PV panel semantic segmentation of satellite and aerial images for better using machine learning," *Advances in Applied Energy*, vol. 4, p. 100057, Nov. 2021, doi: 10.1016/j.adapen.2021.100057.
[24] M. Zech, H.-P. Tetens, and J. Ranalli, "Toward global rooftop PV detection with Deep Active Learning," *Advances in Applied Energy*, p. 100191, Sep. 2024, doi: 10.1016/j.adapen.2024.100191.
[25] R. Pueblas Núñez and L. Kotzur, *FZJ-IEK3-VSA/PASSION*. (Mar. 06, 2024). Jupyter Notebook. FZJ-IEK3. Accessed: May 06, 2024. [Online]. Available: https://github.com/FZJ-IEK3-VSA/PASSION
[26] Z. Zhang et al., "Carbon mitigation potential afforded by rooftop photovoltaic in China," *Nat Commun*, vol. 14, no. 1, p. 2347, Apr. 2023, doi: 10.1038/s41467-023-38079-3.
[27] K. Bódis, I. Kougias, A. Jäger-Waldau, N. Taylor, and S. Szabó, "A high-resolution geospatial assessment of the rooftop solar photovoltaic potential in the European Union," *Renewable and Sustainable Energy Reviews*, vol. 114, p. 109309, Oct. 2019, doi: 10.1016/j.rser.2019.109309.
[28] S. Joshi, S. Mittal, P. Holloway, P. R. Shukla, B. Ó Gallachóir, and J. Glynn, "High resolution global spatiotemporal assessment of rooftop solar photovoltaics potential for renewable electricity generation," *Nat Commun*, vol. 12, no. 1, p. 5738, Oct. 2021, doi: 10.1038/s41467-021-25720-2.
[29] L. Sander, D. Schindler, and C. Jung, "Application of Satellite Data for Estimating Rooftop Solar Photovoltaic Potential," *Remote Sensing*, vol. 16, no. 12, p. 2205, Jun. 2024, doi: 10.3390/rs16122205.
[30] A. Semenuk, "Current Issues, Priority Directions, and Development Strategies of Ukraine," presented at the III International Scientific and Practical Online Conference, ITTA, 2021, p. 1463.
[31] "Statistics Service of Ukraine." [Online]. Available: https://www.ukrstat.gov.ua
[32] "Housing stock of Ukraine." [Online]. Available: https://www.ukrstat.gov.ua/druk/publicat/Arhiv_u/15/Arch_gf_bl.htm
[33] "Open data of the land cadastre of Ukraine." [Online]. Available: https://kadastr.live
[34] T. Pelser, J. M. Weinand, P. Kuckertz, R. McKenna, J. Linssen, and D. Stolten, "Reviewing accuracy & reproducibility of large-scale wind resource assessments," *Advances in Applied Energy*, vol. 13, p. 100158, Feb. 2024, doi: 10.1016/j.adapen.2023.100158.

# Supplementary Material

## S1 Methodology – Factor Extraction

### Correction of Satellite-based Building Footprint Areas

As previously explained, the openly available geospatial datasets for Ukraine do not suffice to directly determine the building footprints. Therefore, East German high-resolution building footprint data of 2021 from the German Federal Office for Cartography and Geodesy [1] was used to derive footprint-area correction factors for building footprints. This need arises from some limitations inherent in the currently most detailed datasets, the WSF [2] and the Microsoft GlobalML [3] datasets, both openly available with at least an almost global coverage. While the WSF contains raster data of built-up areas at a resolution of 10m, Microsoft GlobalML provides building polygons describing each building footprint separately in vector format. Hence, the WSF, due to its format, is unable to properly capture exact building footprints and is therefore unsuitable to directly draw conclusions on the total building footprint area of a region. The total building footprint area is the sum over all footprint areas, i.e., the areas of the polygons describing the building perimeter, of all buildings in the respective region. GlobalML, though able to provide the exact shapes of buildings, lacks coverage in several oblasts in Ukraine like for example around Rivne and Khmel in the North-West, in Chernihiv and Poltava east of Kyiv, in Kherson or along the Carpathian Mountains (see supplementary material) and is thus not suitable as the sole input data source either. Moreover, no official dataset with information on the building footprints is openly available for Ukraine. Therefore, data for East Germany was employed to derive corrected building footprint areas based on footprint-area correction factors. Those factors were then applied to Ukrainian satellite-based footprint data,, based on the building stock similarity [4] as outlined before. GlobalML data was used for regions with full GlobalML coverage and WSF data for the other regions. For Germany, the WSF and GlobalML datasets show the same limitations as for Ukraine. However, an official dataset with actual building footprints [1] is additionally available for Germany. This dataset served to derive the aforementioned footprint-area correction factors $c_{wsf}$ for WSF and $c_{globalml}$ for GlobalML by comparing the total building footprint area per NUTS-3 region in East Germany according to these datasets to that of the official German dataset and averaging over NUTS-3 regions.

Figure 1 provides an exemplary illustration of the GlobalML and WSF data with their respective building footprints compared to the actual ones. As the WSF data provides only binary cell values, the dataset had to be preprocessed by calculating the metric area of each raster cell (yellow squares in Figure 1). The footprint-area correction factors $c_{globalml}$ and $c_{wsf}$ for the building footprints were used to correct the areas of the yellow and lilac boxes to match the area of the buildings outlined in green. In the shown example it is clearly shown that GlobalML has a better match with the actual building footprints.

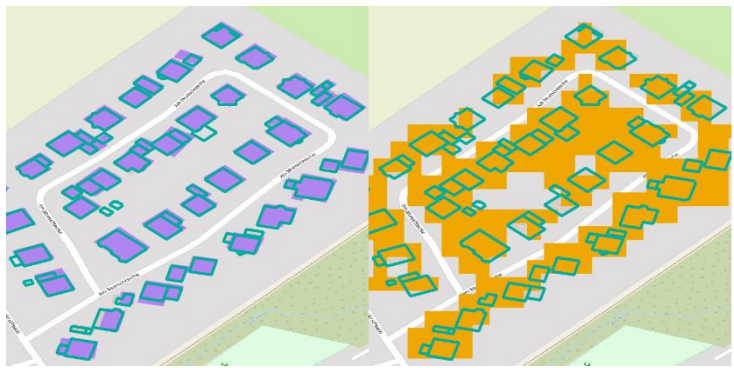

**Figure 1**. Example for GlobalML [3] vector data (left, building footprints in lilac) and WSF [2] raster data (right, built-up area in yellow) in East Germany. The actual building footprints taken from the German Federal Office for Cartography and Geodesy [1] are represented by the green outlines. Background map from OpenStreetMap [5].

The ratios of actual footprint area per NUTS-3 region to footprint area according to the satellite dataset per NUTS-3 region were calculated per NUTS-3 region, grouped by buildings usage types, according to Equation 1, with $n$ as the number of polygons in region.

$$ratio_{region, usage\ type} = \frac{\sum_{i=0}^{n} A_{footprint, usage\ type}}{\sum_{i=0}^{n} A_{footprint, satellite, usage\ type}} \tag{1}$$

These ratios were then aggregated by weighted-averaging over all NUTS-3 regions. However, to add more spatial details to the footprint-area correction factors to account for differences in the morphology of the built-up environment, correction factors were further differentiated between building usage type, i.e., residential and non-residential buildings, and urbanization levels of the NUTS-3 regions in the aggregation process. Residential and non-residential buildings were differentiated by spatially joining the building polygons in all three datasets, the official building footprints, WSF and GlobalML, with the GHS-BUILT-S R2023 dataset for 2018 from the GHSL data package 2023 [6]. This data package was selected due to its global coverage which allows applying it to both the German and the Ukrainian data and ensures the transferability of the workflow to other regions. The GHS-BUILT-S R2023A dataset provides information in raster format at a resolution of 10 m on the location of non-residential built-up areas. All building polygons that touched at least one pixel classified as non-residential were assigned a non-residential usage type. The total footprint areas of residential and non-residential buildings were then calculated for each NUTS-3 region. For the assignment of urbanization levels to NUTS-3 regions we used the Global Degree of Urbanisation Classification of administrative units (R2023) from the same data package. It contains a classification of each administrative unit into urban or rural areas on various administrative levels (GADM). For our analysis, we use the L1 classification [6, p. 62], which has the three levels: (1) rural areas, (2) towns and semi-dense areas, and (3) cities each on GADM level 2 (GID-2). After mapping GID-2 to NUTS-3, each NUTS-3 region was assigned an urbanization level. Then, the final footprint-area correction factor vectors $c_{globalml}$ and $c_{wsf}$ of dimensions 3x2 could be derived by averaging over all NUTS-3 regions grouped by the three levels of urbanization and the two usage types. During averaging, the respective footprint-area correction factors per usage type in the individual NUTS-3 regions were weighted by the area of that usage type, in order to give higher weight to footprint-area correction factors from regions with larger built-up areas. To ensure that outliers as a result of missing data do not influence the mean, values outside of 1.5 times the inter-quartile range were disregarded.

## Translation of Footprint Areas to Roof Areas

Based on the corrected building footprint area, the actual roof area needed to be derived. This data is required for detailed, highly spatially resolved PV potential analyses, as the RTPV potential is directly linked to the roof area. Except for the case of flat roofs, the roof area of a building is not equal to its footprint areas. The roof data was extracted from an internal version of ETHOS.BUILDA[1], the German building database [7], [8]. The underlying datasets for building footprints and roof areas are primarily official 3D building datasets [9] and secondarily OpenStreetMap (OSM) data [5], for the German federal states where the former data was not available at the time of conducting this analysis, which, for East Germany, was only the case for Mecklenburg-Western Pomerania. The 3D building data provides detailed information on the geometry of buildings, making it possible to retrieve not only building footprints but also roof characteristics like the roof areas. OSM data on the other hand does not include detailed roof data. Hence, only those areas with 3D building data were suitable for the calculation of the footprint-to-roof-area translation factors. For those suitable areas within the East German federal states, first, the ratio between the roof area and the building footprint area of each building (footprint-to-roof-area translation factor) was calculated. In addition, we differentiated buildings by usage types as described for the footprint-area correction factor. The building-level footprint-to-roof-area translation factors were averaged per building usage type and degree of urbanization category, weighted by the footprint areas of the buildings. These footprint-to-roof-area translation factors derived from East German data were later applied to the corrected and filtered building footprints of Ukraine.

---

[1] A public version with a reduced set of attributes is available at https://ethos-builda.fz-juelich.de.

### Translation of Roof Areas to RTPV Capacity

Based on the roof areas, the RTPV capacity was derived. Data on the RTPV capacity per roof area is available from Risch, Maier et al. [10]. They calculated RTPV capacity and generation potential for all German buildings based on 3D building data in the Level of Detail 2 (LoD2), which describes a building as a 3-dimensional, rectangular block with a standardized roof shape. Risch, Maier et al., employed a usage factor of 0.6 as done by the IEA [11] for reducing the capacity value to account for shading or constructions on roofs. The ratios were available from the input data at the individual-building level and were then aggregated to statistical values following the same procedure as for the roof-area-to-capacity translation factors $t_r$, which serve as input for the further analysis.

### Roof Tilt and Orientation Statistics

Finally, statistics regarding the share of capacity per roof tilt and azimuth were calculated from the East German building data. This was achieved by retrieving the roof areas per tilt and azimuth category for each building type and degree of urbanization category and multiplying the roof areas with the previously calculated roof-area-to-capacity translation factor. The tilt and azimuth categories are presented in Table 1. The shares of the total capacity per tilt and azimuth combination were then calculated. The roof area as well as the maximum installable rooftop capacity per polygon was then calculated using the corrected footprint area and the two translation factors determined above, depending on usage type and degree of urbanization.

**Table 1**. Tilt and azimuth buckets in roof statistics analysis.

| Characteristic | Buckets |
| --- | --- |
| Azimuth | North, East, South, West |
| Tilt | 0-10, 10-20, 20-30, 30-40, 40-50, 50-60, >60 |

# S2 Methodology – Ignored Polygons

The following list contains all building types that were deemed irrelevant and were thus ignored for the calculation of the RTPV potential in Ukraine.

'25', '27', '54', '81', 'abandoned', 'abandonedq', 'ah', 'allotment_house', 'antenna', 'bandstand', 'barrier', 'basilica', 'bell_tower', 'bird_cage', 'block', 'boathouse', 'bomb_shelters', 'bridge', 'bunker', 'bus_station', 'bus_stop', 'canopy', 'castle', 'castle_wall', 'cathedral', 'cellar', 'changing_rooms', 'chapel', 'checkpoint', 'chimney', 'church', 'collapsed', 'conservatory', 'construction', 'container', 'cooling_tower', 'corridor', 'cross_box', 'crossing_box', 'demolitions', 'digester', 'disuse', 'disused', 'dogshed', 'ed', 'enclosures', 'fence', 'floating_home', 'font', 'foo', 'footway', 'forestry', 'forsaken', 'garden_house', 'gasometer', 'gatehouse', 'gazebo', 'glasshouse', 'goatshed', 'grandstand', 'granstand', 'grass', 'greenhouse', 'greenhouse_horticult', 'halt', 'heath', 'heritage', 'historic', 'historical', 'history', 'houseboat', 'hut', 'kingdom_hall', 'kiosk', 'map', 'military', 'mill', 'minor_distribution', 'monitoring_station', 'monument,_war_memori', 'mosque', 'palace', 'parking', 'parking_entrance', 'part', 'passageway', 'pavilion', 'presbytery', 'propane_tank', 'proposed', 'pumping_station', 'quonset_hut', 'railcar', 'razed', 'red', 'religious', 'reservoir', 'ruine', 'ruins', 'rк', 's',

'security_post', 'shed', 'shelter', 'ship', 'shrine', 'silo', 'site', 'stadium', 'stand', 'stands', 'static_caravan', 'stilt_house', 'storage_tank', 'synagogue', 'tank', 'temple', 'tent', 'terrace', 'terraced', 'toilet', 'toilets', 'tomb', 'tool', 'tower', 'trai', 'transformer_tower', 'transportation', 'tree_house', 'tribune', 'underconstructing', 'wall', 'wayside_shrine', 'wc', 'windmill', 'yes;construction', 'yfence', 'н']

# S3 Result – Factor Extraction

## Correction and Translation Factors from East German Data

Figure 2 shows the footprint-area correction factors for GlobalML and WSF data, which were calculated based on East German data as described in S1. The factors for GlobalML are closer to 1 than those for WSF. This indicates that the footprint areas in GlobalML are more similar to the actual footprint areas than the areas in WSF, as 1 would equal a perfect match. For the WSF dataset, the footprint-area-correction factors remain below 1, which means that it overestimates the footprint areas. The footprint-area correction factors for GlobalML lie within a range of 0.86 to 1.07 across all urbanization categories and usage type categories. Means increase from rural areas over town & semi-dense areas to cities, which shows that the alignment is highest in cities, whereas the overestimation of building areas is most pronounced in rural areas. As previously mentioned, the footprint-area correction factors for the WSF data are lower and lie within a range of between 0.37 and 0.57. This is attributable to the fact that the WSF data is raster data at a resolution of 10x10m. Therefore, it cannot capture the exact extent of individual buildings. Instead, an area with smaller buildings might be represented by a continuous area of pixels indicating built-up area, therefore resulting in larger built-up areas when summing up the area covered by the pixels. The separate lower footprint-area correction factor counteracts this classification inaccuracy to yield realistic footprint areas, independent of the data source.

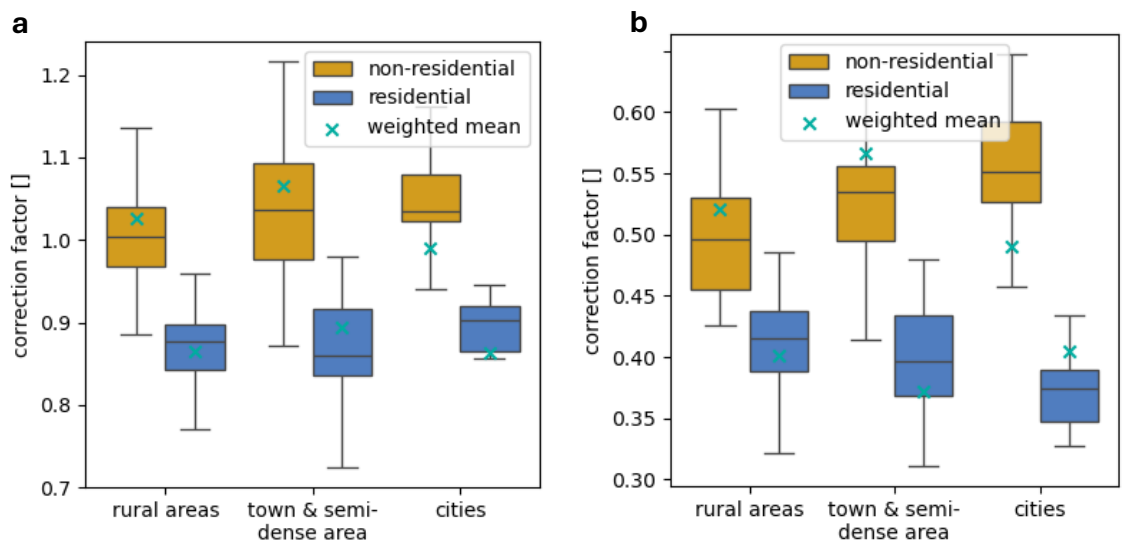

**Figure 2.** Footprint-area correction factors calculated based on East German data grouped by DUC and usage type without outliers for a) GlobalML and b) WSF.

Apart from the footprint-area correction factors, factors for determining the roof area and the RTPV capacity from the footprint area were calculated. These factors are illustrated in Figure 3, grouped by building usage type and DUC. The average footprint-to-roof-area translation factors lie between 1.08 and 1.23, the total roof area is hence 8-23% larger than the total footprint area. In the case of flat roofs, the factor would be 1, i.e., equal to the footprint area. In residential buildings, the factor is higher, which is attributable to the higher share of non-flat roofs, such as gabled roofs, for this usage type. The same explanation is also applicable to lower factor values in cities. Higher shares of apartment buildings and multi-family houses entail larger shares of flat roofs, which result in a lower footprint-to-roof-area translation factor. The factor for deriving RTPV capacity from roof areas falls in the range of 0.08 kW/m² to 0.1 kW/m².

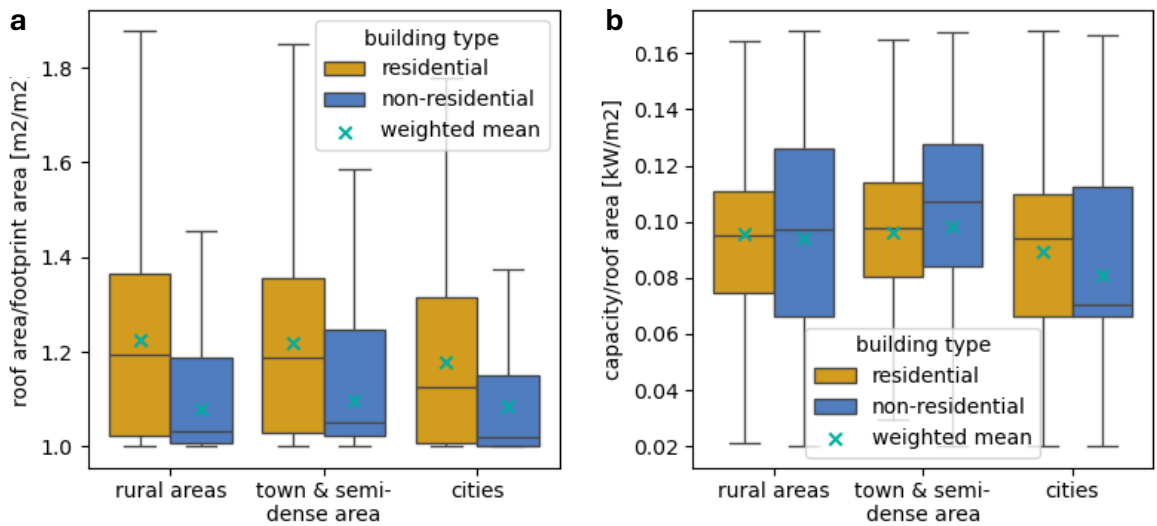

**Figure 3**. Factors for determining a) footprint-to-roof-area translation factor and b) the roof-area-to-capacity translation factor from East German 3D building data. Outliers are not displayed.

Finally, the distribution of roof areas by azimuth and tilt was determined for East German buildings and is depicted in Figure 4. Figure 4 a) shows that for non-residential buildings, flat roofs with a tilt of 0-10° dominate with a share of 55.9%, followed by roofs with angles of 10-20° with a share of 25.9%. For residential buildings, flat roofs and roofs with a tilt of 40-50° are most prevalent with shares of 29.6% and 25.1%, respectively. Figure 4 b) shows that the orientation of roof areas is nearly independent of the building type, differences in the shares of the orientations between residential and non-residential remain below 1.5 percentage points. Furthermore, roof areas are almost equally distributed between the four orientation categories, with North and South orientation together 3.0 and 7.5 percentage points more frequent than East and West orientation for residential and non-residential buildings, respectively.

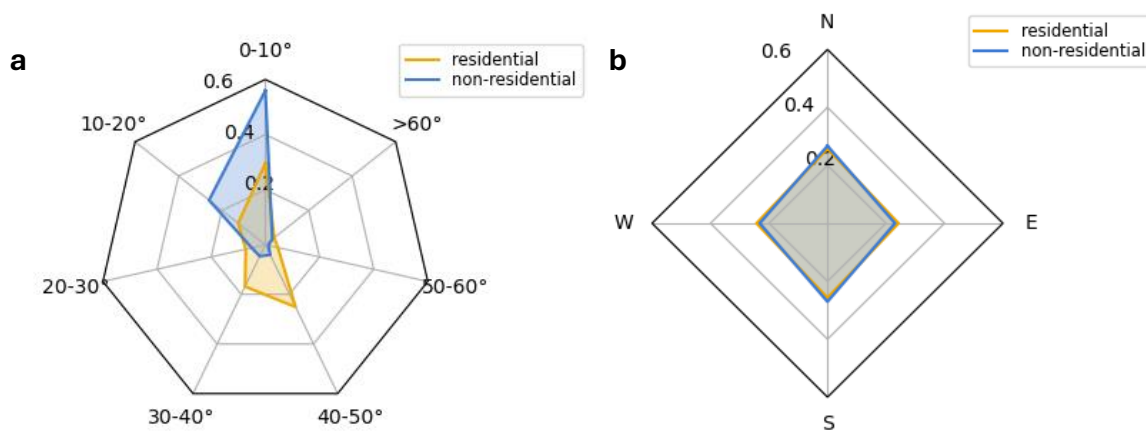

**Figure 4**. Distribution of roof areas in East German buildings by a) tilt and b) orientation (excluding flat roofs with a tilt of 0-10°).

# S4 Results – RTPV Capacity and Generation Potential

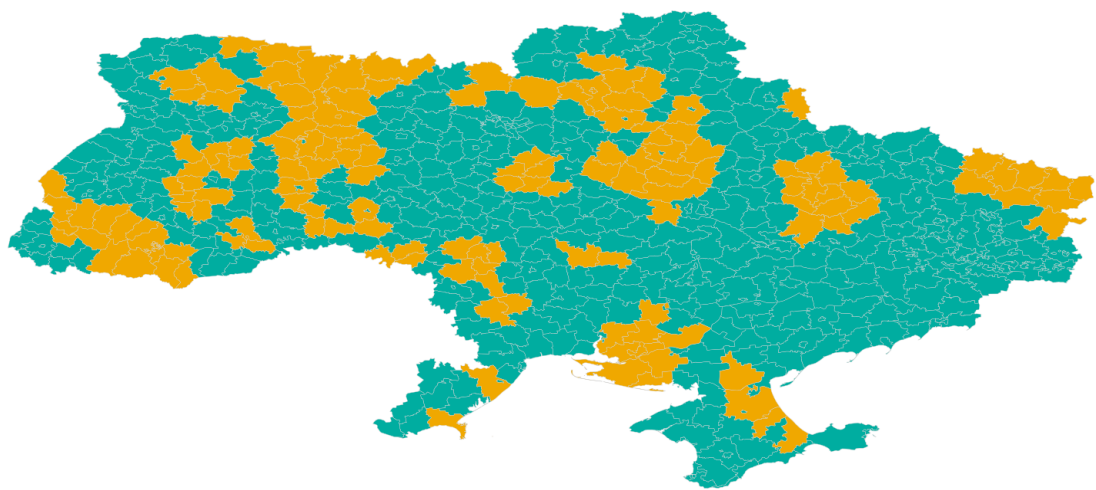

**Figure 5**. Building footprint data source per district; green indicates the use of GlobalML vector data and orange the use of World Settlement Footprint 2019 raster data.

Table 2. RTPV capacity potential per oblast and azimuth.

| Oblast | Capacity [GW] | | | | | | Capacity density [MW/km²] |
|---|---|---|---|---|---|---|---|
| | flat | South | East | West | North | total | |
| Cherkasy | 3.0 | 1.4 | 1.3 | 1.3 | 1.4 | 8.4 | 400 |
| Chernihiv | 2.7 | 1.5 | 1.4 | 1.4 | 1.5 | 8.5 | 261 |
| Chernivtsi | 2.3 | 1.3 | 1.2 | 1.2 | 1.3 | 7.4 | 897 |
| Crimea | 4.7 | 2.4 | 2.2 | 2.2 | 2.4 | 13.9 | 528 |
| Dnipropetrovs'k | 7.1 | 3.0 | 2.8 | 2.8 | 3.0 | 18.7 | 589 |
| Donets'k | 9.0 | 3.3 | 3.2 | 3.2 | 3.3 | 22.0 | 827 |
| Ivano-Frankivs'k | 2.6 | 1.5 | 1.4 | 1.4 | 1.5 | 8.3 | 603 |
| Kharkiv | 5.4 | 2.5 | 2.4 | 2.4 | 2.5 | 15.2 | 484 |
| Kherson | 4.2 | 2.2 | 2.0 | 2.0 | 2.2 | 12.5 | 490 |
| Khmel'nyts'kyy | 2.8 | 1.5 | 1.4 | 1.4 | 1.5 | 8.5 | 411 |
| Kyiv | 5.3 | 3.0 | 2.8 | 2.8 | 3.0 | 16.8 | 596 |
| Kyiv City | 2.8 | 0.6 | 0.6 | 0.6 | 0.6 | 5.3 | 6351 |
| Kirovohrad | 2.6 | 1.2 | 1.1 | 1.1 | 1.2 | 7.2 | 289 |
| L'viv | 4.7 | 2.4 | 2.3 | 2.3 | 2.4 | 14.1 | 646 |
| Luhans'k | 4.9 | 1.9 | 1.8 | 1.8 | 1.9 | 12.4 | 460 |
| Mykolayiv | 3.0 | 1.6 | 1.5 | 1.5 | 1.6 | 9.1 | 383 |
| Odesa | 5.6 | 2.7 | 2.6 | 2.6 | 2.7 | 16.2 | 485 |
| Poltava | 3.4 | 1.7 | 1.6 | 1.6 | 1.7 | 10.1 | 350 |
| Rivne | 2.4 | 1.4 | 1.3 | 1.3 | 1.4 | 7.7 | 384 |
| Sevastopol' City | 0.7 | 0.2 | 0.2 | 0.2 | 0.2 | 1.6 | 2004 |

| Sumy | 2.7 | 1.4 | 1.3 | 1.3 | 1.4 | 8.2 | 347 |
| Ternopil' | 2.1 | 1.2 | 1.1 | 1.1 | 1.2 | 6.7 | 484 |
| Transcarpathia | 2.6 | 1.5 | 1.4 | 1.4 | 1.5 | 8.4 | 655 |
| Vinnytsya | 3.6 | 2.1 | 2.0 | 2.0 | 2.1 | 11.8 | 445 |
| Volyn | 2.4 | 1.2 | 1.2 | 1.2 | 1.3 | 7.3 | 357 |
| Zaporizhzhya | 4.5 | 1.9 | 1.8 | 1.8 | 1.9 | 12.1 | 452 |
| Zhytomyr | 2.9 | 1.5 | 1.4 | 1.4 | 1.5 | 8.7 | 289 |
| Ukraine | 100.0 | 48.1 | 45.3 | 45.3 | 48.2 | 287.0 | 477 |

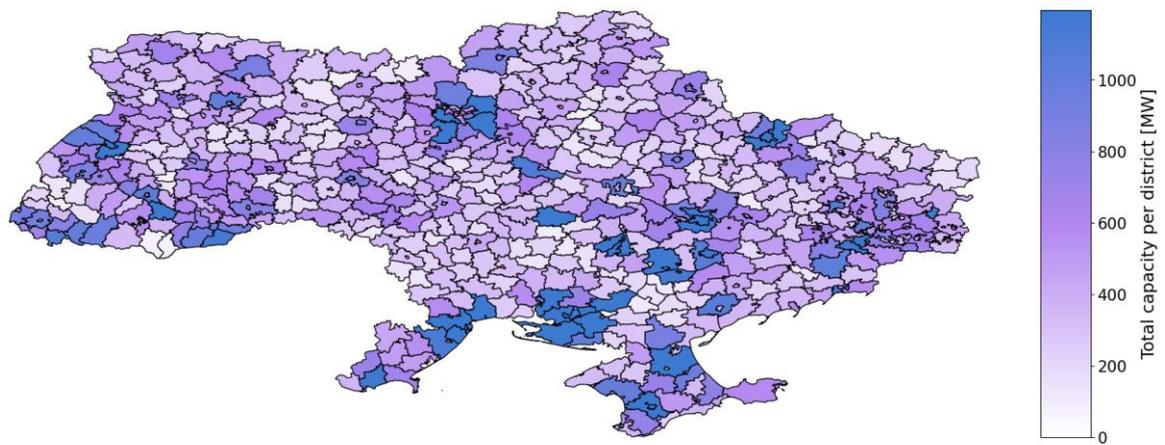

**Figure 6**. District level capacities, clipped at the 95th percentile.

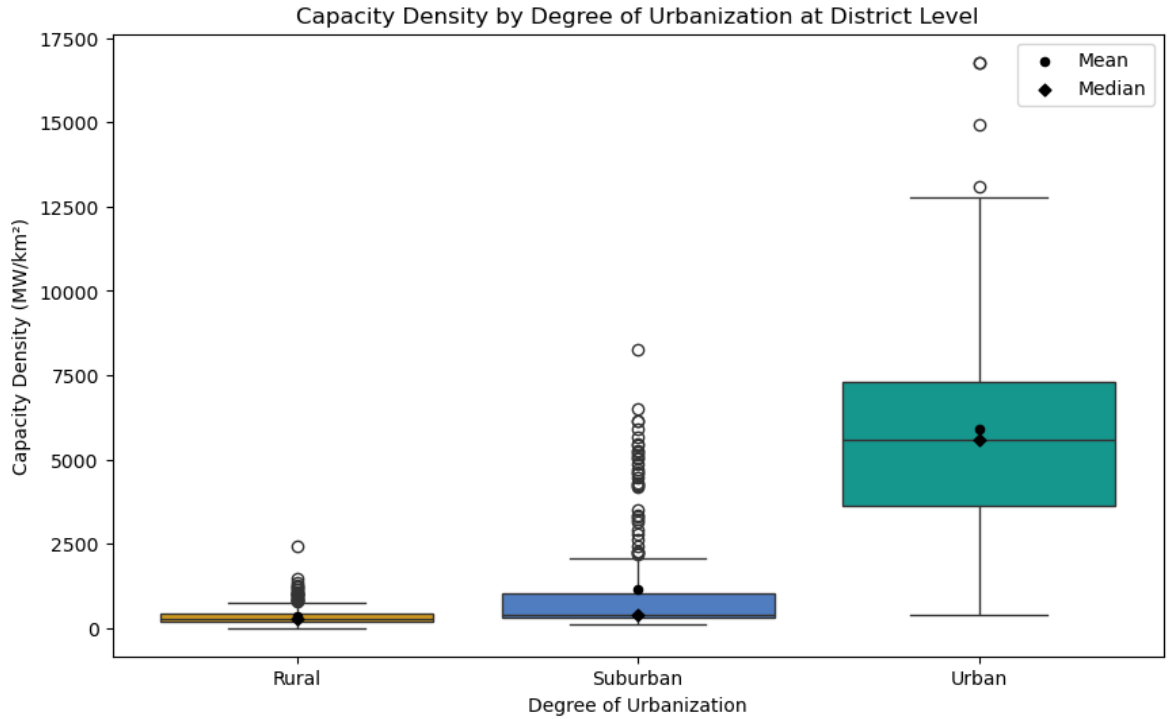

**Figure 7**. Capacity density by degree of urbanization at district level.

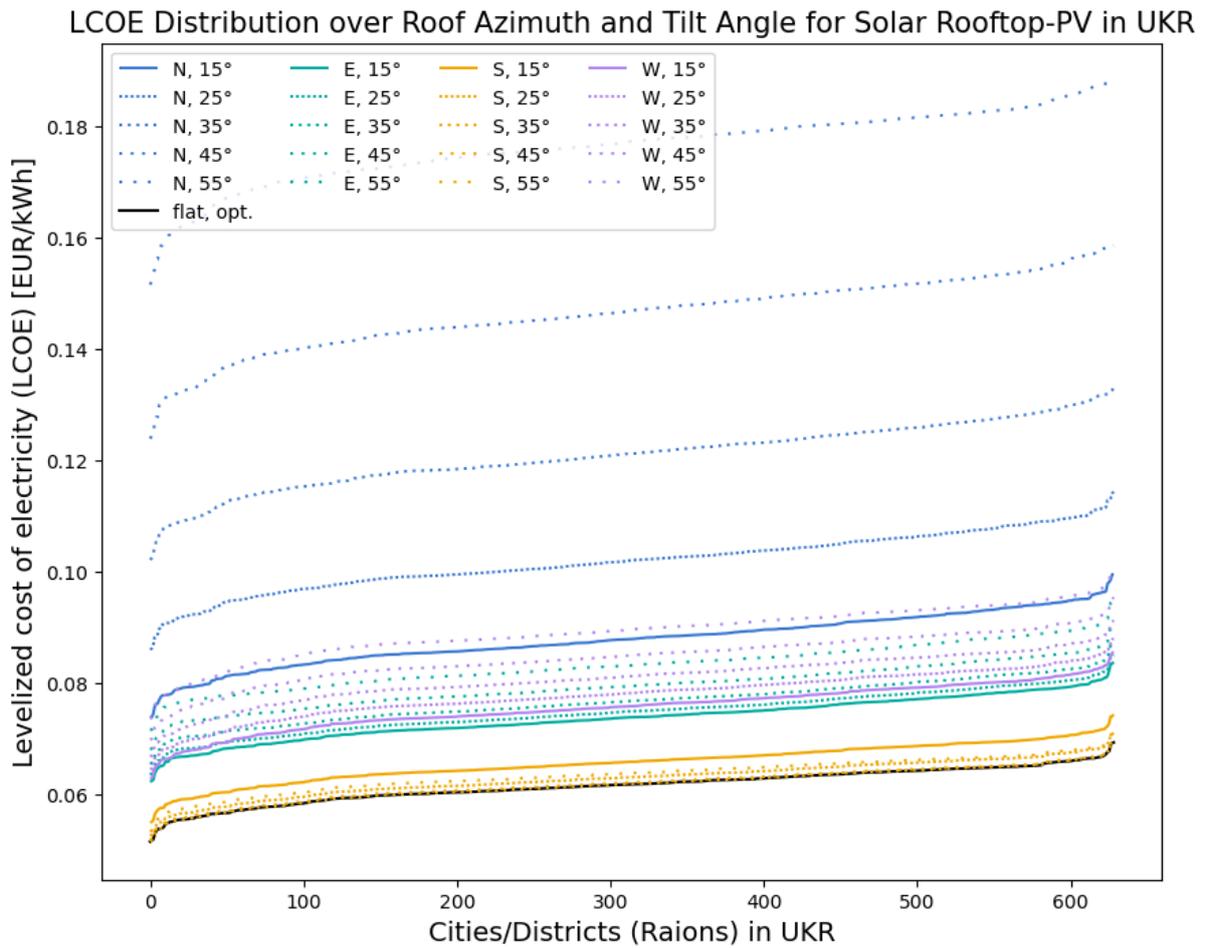

**Figure 8**. Impact of location (district) and module orientation onto levelized cost of electricity.

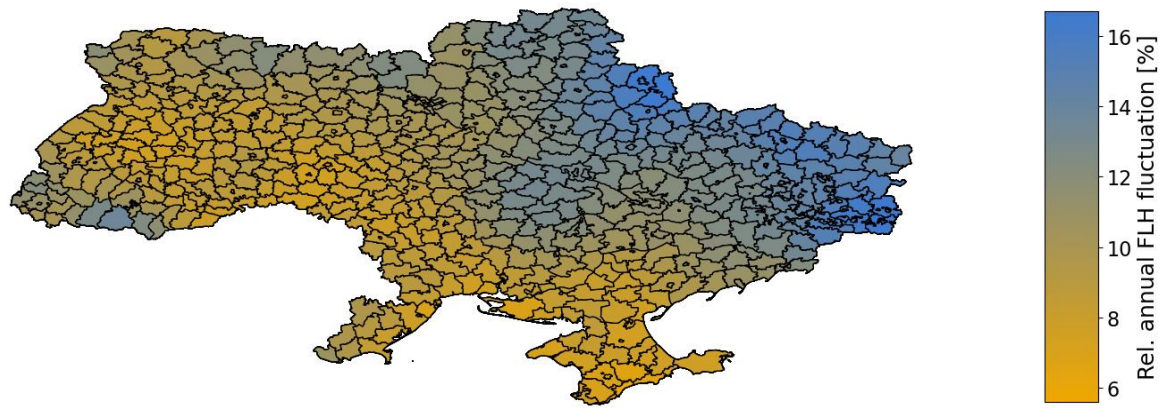

**Figure 9**. Relative fluctuation between annual FLH for systems with optimal orientation (2000-2019).

## S5 Validation

In the plausibility checks of the results, some assumptions were made in cases where direct calculations were not available:

**Footprint area of private houses in Ukraine:**

- *In large cities:* 154 m²  [12].

- *In small towns:* 70 m² [13].

**Assumption of average building numbers per private yards in Ukraine:**

- *In large cities:* 2 buildings

- *In small towns/villages:* 3.5 buildings (Main building, summer kitchen, livestock building, a little tool warehouse  or garage)

**Distribution of building types in Ukraine:**

- *Apartment buildings:* Buildings owned by the state, usually multi-family houses/high-rise buildings. In our publication it belongs to the type of residential building.

- *Private households:* Private houses, usually single-family houses. In our publication it belongs to the type of residential building.

- *Municipal buildings:* Schools, kindergartens, etc. In our publication it belongs to the type of non-residential building.

- *Industry:* Old Soviet production factories, other companies/manufacturers. In our publication it belongs to the type of non-residential building.

First, a detailed analysis of various Ukrainian cities was conducted to assess the built-up areas. This included reviewing publications, consulting with experts, and analyzing policy documents to gather data on the roof areas of various building types. Personal contacts and published reports were also used to obtain relevant roof information.

Then, when data was not available for specific cities or regions, estimates were made by combining data from multiple sources. When information was not available for a specific building type, estimates were made using averages derived from similar buildings in both smaller and larger cities.

In addition, the potential rooftop electricity generation in kilowatts for solar installations was estimated by analyzing various reports that provide data on theoretical maximum capacity for different building types and cities. These results were then compared with Ukrainian potential studies as well as global studies to assess their plausibility and consistency. The estimates were cross-checked with studies focusing on individual buildings, entire cities, or specific building types.

According to the Information and Analysis Note "Model Scenario Assessments for Zhytomyr's Transition to 100% Renewable Energy Sources by 2050", the total approximate footprint area of the city is 5,293,200 m² [14]. Using the Public Cadastral Map of Ukraine and its built-in area measurement tool, the roof areas of large industrial buildings were calculated to be approximately 650,000 m². Another 100,000 m² belong to municipal buildings, schools and kindergartens. The city of Zhytomyr has 2,258 residential buildings. By measuring the roof areas of a 10% sample, the average roof area of such a building was determined to be 1,400 m². It is also assumed that 30% of all roofs are inclined, so that only half of their area is suitable for the installation of rooftop solar power plants. Based on these assumptions, the total roof area of residential buildings where rooftop PV can be installed is approximately 2,700,000 m². There are 13,152 private houses in Zhytomyr, so the suitable footprint areas are 1,972,800 m² for private houses. The coefficient between built-up area and roof area of our calculation for the city of Zhytomyr is about 0.84. Taking into account the roofs of 30% of inclined buildings, the area of the residential blocks is 2,570,400 m². Otherwise, it is assumed that the roof area is equal to the footprint area due to the fact that the municipal buildings, industrial buildings and residential blocks in Ukraine usually have non-sloping roofs. Our publication shows that the total footprint area of the city of Zhytomyr is 7,797,060 m².

Based on information from the Poltava city portal and e-mail communication with the city authorities, it is estimated that the total built-up area of Poltava is 5,120,766 m² , of which 655,000 m² is the assumed industrial zone [14]. The total built-up area of private houses is 2,922,766 m², and the area of apartment buildings together with municipal buildings is 1,543,000 m² [15], [16].In comparison, our study shows that the total built-up area of the city is about 9,674,784 m².

The relatively large errors in the footprint areas for the cities of Poltava and Zhytomyr could be attributed to the way authorities define the number of private houses in publicly available information. It remains unclear whether non-residential private buildings within the yards are included in these figures.

As stated in the "Sustainable Energy and Climate Action Plan for the City of Enerhodar until 2030", the city itself has 244 apartment buildings with a total roof area of 236,625 m². The roof area of the municipal buildings in Enerhodar is 92,000 m², so the total roof area of the city is 328,625 m² [17]. The calculations for Enerhodar do not include private households because the city was planned for the nuclear power plant and most of the buildings are apartment blocks. In addition, the only two industrial complexes are the nuclear power plant and the heating plant, which are not suitable for rooftop PV.

According to the "Sustainable Energy and Climate Action Plan until 2030 in Novogrodivka" [13], the calculated built-up area of the city is 531,000 m², of which 141,000 m² are municipal residential buildings, 45,550 m² are municipal non-residential buildings, and 295,000 m² are private residential buildings (98,420 m² + supplementary buildings consideration). The calculation of municipal residential buildings was estimated using the available information on the number of different types of buildings, with the total area of apartments and the distribution of floors (taking into the consideration weighted average stories coefficient of 2.68: 261,300 / 2.6875 = 97,250 m²). It should also be noted that 43,750 m² (30%) [18] of municipal residential buildings (97,250 + 43,750 = 141,00 m²) were added to the calculation due to the consideration of walls, which are part of the footprint area but not part of the living area. According to the results of our study, the built-up area of the town of Novogrodivka is about 560,500 sq. m. It should be noted that our study also includes the areas of industrial buildings, which are not mentioned in the literature.

One of the buildings in the city of Chortkiv was thoroughly studied in the "Analysis of renewable energy potential in the city of Chortkiv and Chortkiv district". For the roof of Chortkiv secondary school No. 5 a solar energy system was modeled using cartographic and meteorological data. This model is approximate and does not take into account all structural elements of the roof. The modeling also assumed that the entire roof is flat, although a part of it has a slight slope. The developed model specifies a panel tilt angle of 35°, facing south. With such an arrangement of modules, the installed capacity of the station is modeled to be 202 kW [19]. The results of our work calculated the rooftop potential to be as much as 383 kW.

Furthermore, the rooftop areas of seven additional municipal buildings were measured, totaling 12,194 m². Our research identified a rooftop area of 11,363 m² for these buildings [19].

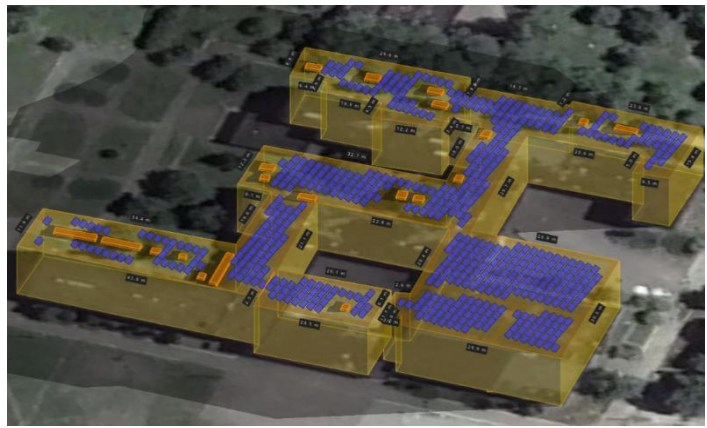

**Figure 10**. Modelling of solar energy system for the roof of Chortkiv Secondary School No. 5, taking into account some special features of the roof.

Table 3 gives a summary of the plausibility assessment of the data from our publication compared with the literature findings for Ukraine. Table 4 shows the calculated deviations between the data of our publication and the references.

**Table 3**. Plausibility check of the data from our publication and the literature research on Ukraine.

| Considered object | Footprint area our work (m²) | Rooftop area our work (m²) | Capacity our work (GW) | Footprint area literature (m²) | Rooftop area literature (m²) | Capacity literature (GW) |
|---|---|---|---|---|---|---|
| Ukraine | - | - | 239 | - | - | 233.6 |
| Zhytomyr | 7,797,060 | - | 846 | 5,293,000 | - | 508 |
| Poltava | 9,674,784 | - | - | 5,120,700 | - | - |
| Enerhodar | - | 292,980 | 74 | - | 328,625 | 86 |
| Novogrodivka | 560,500 | - | - | 531,000 | - | - |
| Ladyzhyn (18 buildings) | - | - | 0.00165 | - | - | 0.0039 |
| Chortkiv (School No.5) | - | - | 0.000383 | - | - | 0.0002 |
| Chortkiv (7 buildings) | - | 11,363 | - | - | 12,194 | - |

Table 4. Deviation between the data from this study and the reference data.

| Considered object | Footprint area error (%) | Rooftop area error (%) | Capacity error (%) |
|---|---|---|---|
| Ukraine | - | - | 2.5 (residential only) |
| Zhytomyr | 33 | - | 40 |
| Poltava | 53 | - | - |
| Enerhodar | - | 10.8 | 14 |
| Novogrodivka | 5.6 | - | - |
| Ladyzhyn (18 buildings) | - | - | 58.7 |
| Chortkiv (School No.5) | - | - | 91.5 |
| Chortkiv (7 buildings) | - | 6.8 | - |

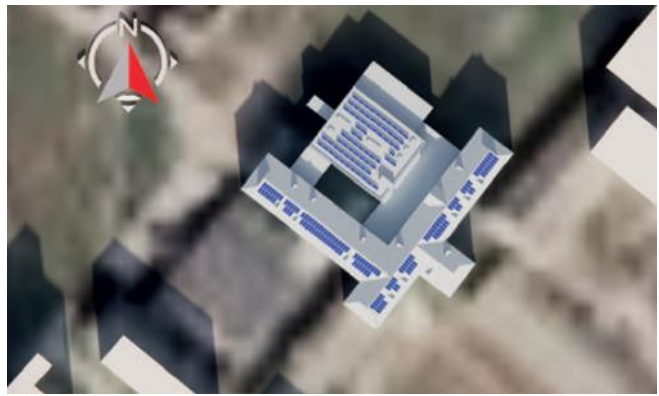

**Figure 11**. Example of modelling the maximum roof potential for the roof of Ladyzhyn School No. 4, taking into account the special features of the roof [14].

# Bibliography

[1] Bundesamt für Kartographie und Geodäsie, "Amtliche Hausumringe Deutschland." Accessed: May 13, 2024. [Online]. Available: https://gdz.bkg.bund.de/index.php/default/digitale-geodaten/sonstige-geodaten/amtliche-hausumringe-deutschland-hu-de.html

[2] "EOC Geoservice Maps - World Settlement Footprint (WSF) - Sentinel-1/Sentinel-2 - Global, 2019." Accessed: Oct. 21, 2024. [Online]. Available: https://geoservice.dlr.de/web/maps/eoc:wsf2019

[3] *microsoft/GlobalMLBuildingFootprints*. (Aug. 09, 2023). Python. Microsoft. Accessed: Aug. 09, 2023. [Online]. Available: https://github.com/microsoft/GlobalMLBuildingFootprints

[4] C. Klusemann, "Dreimal „Platte" – Einblicke in Techniken und Strukturen des Bauens in der DDR," *Kunstchronik. Monatsschrift für Kunstwissenschaft*, pp. 19-28 Seiten, Jan. 2024, doi: 10.11588/KC.2024.1.100239.

[5] OpenStreetMap contributors, *OpenStreetMap*. (2015). [Online]. Available: https://www.openstreetmap.org

[6] European Commission. Joint Research Centre., *GHSL data package 2023.* LU: Publications Office, 2023. Accessed: Mar. 21, 2024. [Online]. Available: https://data.europa.eu/doi/10.2760/098587

[7]     K. Dabrock, N. Pflugradt, J. M. Weinand, and D. Stolten, "Leveraging machine learning to generate a unified and complete building height dataset for Germany," *Energy and AI*, vol. 17, p. 100408, Sep. 2024, doi: 10.1016/j.egyai.2024.100408.

[8]     K. Dabrock, J. Ulken, N. Pflugradt, J. M. Weinand, and D. Stolten, "Generating a Nationwide Residential Building Types Dataset Using Machine Learning," Jul. 03, 2024, *Rochester, NY*: 4884155. doi: 10.2139/ssrn.4884155.

[9]     © GeoBasis-DE / BKG, "3D Gebäudemodell LoD2 Deutschland." 2021. [Online]. Available: https://gdz.bkg.bund.de/index.php/default/3d-gebaudemodelle-lod2-deutschland-lod2-de.html

[10]    S. Risch *et al.*, "Potentials of Renewable Energy Sources in Germany and the Influence of Land Use Datasets," *Energies*, vol. 15, no. 15, p. 5536, Jan. 2022, doi: 10.3390/en15155536.

[11]    International Energy Agency (IEA), "Potential for Building Integrated Photovoltaics," International Energy Agency (IEA), Paris, IEA-PVPS T7-4 : 2002, 2002.

[12]    "Середня площа нових квартир в Україні 59 кв. м | РБК Украина." Accessed: Nov. 28, 2024. [Online]. Available: https://www.rbc.ua/ukr/news/gosstat-nazval-srednyuyu-ploshchad-novyh-1574778929.html

[13]    Novohrodivka City Council, "Sustainable Energy and Climate Action Plan until 2030 in Novogrodivka."

[14]    Zhytomyr City Council, "Information-Analytical Note 'Model Scenario Assessments of Zhytomyr's Transition to 100% Renewable Energy Sources by 2050.'"

[15]    Syrota, "Poltava City Council Email."

[16]    "Житловий фонд." Accessed: Oct. 07, 2024. [Online]. Available: https://open.rada-poltava.gov.ua/statistika/rozdil/35/Zhitloviy-fond

[17]    Enerhodar City Council, "Sustainable Energy and Climate Action Plan of the City of Enerhodar until 2030.," *European Covenant of Mayors Community*.

[18]    D. Bansal, V. K. Minocha, A. Kaur, V. A. Dakwale, and R. V. Ralegaonkar, "Reduction of Embodied Energy and Construction Cost of Affordable Houses through Efficient Architectural Design: A Case Study in Indian Scenario," *Advances in Civil Engineering*, vol. 2021, no. 1, p. 5693101, Jan. 2021, doi: 10.1155/2021/5693101.

[19]    Н. Хоодова, "Analysis of the Renewable Energy Potential of the City of Chortkiv and Chortkiv District".